\documentclass[%
 amsmath,amssymb,
preprint,%
onecolumn,
groupedaddress%
]{revtex4-1}
\usepackage{graphicx}
\usepackage{dcolumn}
\usepackage{bm}

\begin{document}

\preprint{AIP/123-QED}

\title[Large scale flows in transitional plane Couette flow: a key ingredient of the spot growth mechanism]{Large scale flows in transitional plane Couette flow: \\ a key ingredient of the spot growth mechanism}

\author{M. Couliou}
\email{couliou@ensta.fr}
\author{R. Monchaux}%
 \email{monchaux@ensta.fr}
\affiliation{ Unit\'e de m\'ecanique, ENSTA-ParisTech, 828 Boulevard des Mar\'echaux, 91762 Palaiseau Cedex, France
}%

\date{\today}

\begin{abstract}
Using Particle Image Velocimetry (PIV) in a new experimental plane Couette flow, we investigate the dynamics of turbulent patches invading formerly laminar flows. We evidence experimentally for the first time in this geometry the existence of large scale flows. These flows appear as soon as laminar and turbulent domains coexist. Spectral analysis is used to study the dynamical evolution of these large scales as well as that of the small scales associated with turbulence. We show that large-scale flows grow {\it before} turbulent spots develop and we point out the crucial role they play in the growth mechanism and possibly also in the emergence of organised patterns.
 \end{abstract}                             
\maketitle
\section{\label{sec:intro}Introduction}

Transition to turbulence may occur through gradual supercritical scenarios as in Rayleigh-B\'enard convection or in Taylor-Couette flow when the outer cylinder is at rest. It is then well understood in terms of linear stability analysis. Beyond these cases, the transition to turbulence may also happen in a more sudden fashion that linear stability analysis fails to predict. Most wall-bounded shear flows fall into this last category and experience a subcritical transition to turbulence involving the abrupt appearance of a disorganised state within the laminar phase. Plane Couette (PCF), Hagen-Poiseuille or plane Poiseuille flows are typical examples of such a transition. The Reynolds number defined as $Re=UL/\nu$, where $U$ and $L$ are typical velocity and length scales and $\nu$ is the fluid kinematic viscosity, being the natural control parameter of these systems, several critical values of $Re$ can be pointed out. Following linear stability analysis, these flows should become turbulent beyond $Re=Re_c$ but experiments or numerical simulations show that turbulent patches can be sustained for values of $Re$ exceeding some value $Re_g$ with $Re_g<Re_c$ which is a global stability threshold below which any perturbation relaxes to the laminar state. In the PCF case, one can also define $Re_t$ beyond which the flow is always observed to be homogeneously turbulent and below which turbulence only fills a fraction of space $F_t$, hereafter called the turbulent fraction. It follows that a transitional range exists between $Re_g$ and $Re_t$ where laminar and turbulent domains coexist, where the flow is sensitive to finite size/amplitude perturbations and displays a strong hysteresis.\\
In the nineties, two teams in Saclay and Stockholm have studied the transition to turbulence in PCF experimentally, pointing out its subcritical nature\cite{daviaud92_PRL,tillmark92_JFM} and characterising the growth of triggered turbulent patches thus denominated as spots\cite{dauchot95b_POF,tillmark92_JFM}. These turbulent spots consist of localised coherent structures made of small-scale longitudinal vortices and velocity streaks characteristic of wall turbulence and already evidenced and studied in boundary layer\cite{gadelhak81_JFM}, channel flows\cite{carlson82_JFM} and numerical plane Couette flow in the pioneering work of Lundbladh and Johansson\cite{lundbladh91_JFM}. An other striking feature of PCF in the transitional range is the existence of a steady pattern of alternating turbulent and laminar stripes inclined with respect to the mean flow. These patterns were first observed experimentally by Coles and van Atta in Taylor-Couette flow\cite{coles65_JFM,vanatta66_JFM} and by Prigent {\it et al.} in PCF\cite{prigent02_PRL} and since then reproduced numerically by different teams in PCF\cite{barkley05_PRL,duguet10_JFM,rolland11_EPJB} and channel flow\cite{tsukahara10_IUTAM}.\\
Spot growth has been carefully studied by the different teams cited above with similar results: growth rates in the streamwise and spanwise directions have been found to be almost constant during the growth and to be increasing functions of the Reynolds number; waves at the spots' edges have been observed to move at a speed slower than the spreading of turbulence. Regarding the growth mechanism, it could be due to stochastic nucleation of new streaks at the edge of the spot associated with a local destabilisation of the velocity profile at the laminar-turbulent interface, or to non-normal growth. A seemingly key element already pointed out in the work of Lundbladh and Johansson\cite{lundbladh91_JFM} is the large-scale flow that develops around the turbulent growing spots with a kind of quadrupolar structure that has been observed numerically in PCF\cite{lundbladh91_JFM,duguet13_PRL} and in model flows\cite{schumacher01_PRE,lagha07_POF} but which is also present along the regular laminar/turbulent patterns\cite{duguet13_PRL}. So far, they have never been observed experimentally in PCF and only very recently in plane Poiseuille\cite{lemoult13_JFM} in spite of the important role they may play regarding spot growth and pattern sustainment as recently suggested by Duguet and Schlatter\cite{duguet13_PRL} who proposed a mechanism combining stochastic nucleation of new streaks at the spot edges and their advection by large scale flows to explain the spot growth and its tendency to form oblique patterns simultaneously.\\
In the present article, we aim at evidencing large-scale flows at the laminar-turbulent interfaces in an experimental plane Couette flow and at quantifying them as much as possible all along the growth process. This would be a first step in the precise understanding of turbulence spreading in the transitional range of PCF. The article is organised as follows: we first describe our experimental set-up, the acquisition chain and the post-processing methods in \ref{sec:meth}. In section \ref{sec:res}, dedicated to quantitative measurements of large-scale flows in two very different laminar-turbulent coexistence situations, we particularly insist on the dynamical aspects. Implications of these results on spot growth mechanisms and also possibly on the onset of organised patterns are further discussed in \ref{sec:disc}.
\section{\label{sec:meth}Experimental set-up \& Methods}
\subsection{Experimental set-up and measurement systems}
\begin{figure}[t!]
\begin{minipage}[h]{.43\linewidth}
   \includegraphics[width=0.95\textwidth]{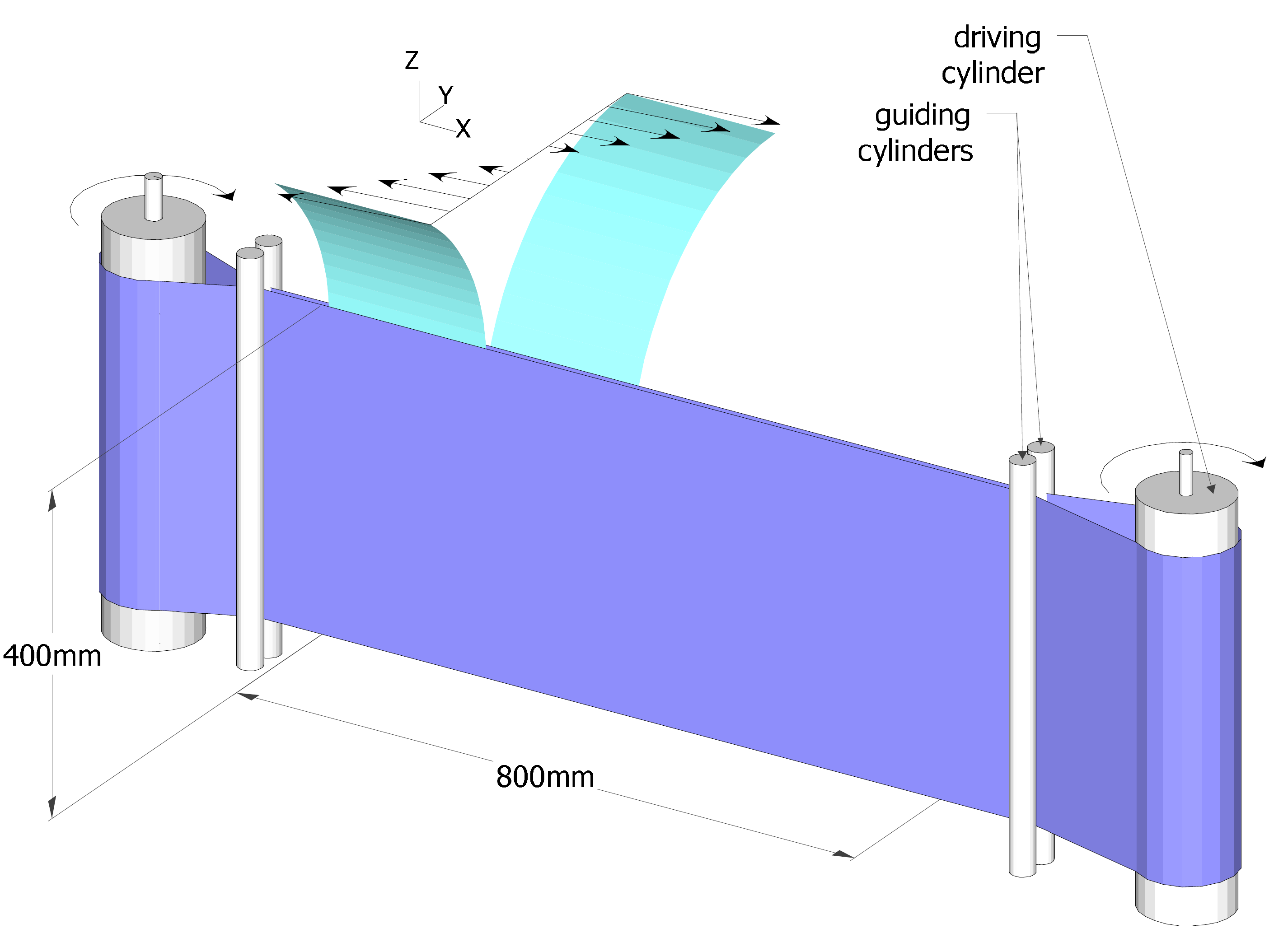}\\
   (a)
\end{minipage}
\begin{minipage}[h]{.47\linewidth}
   \includegraphics[width=0.93\textwidth]{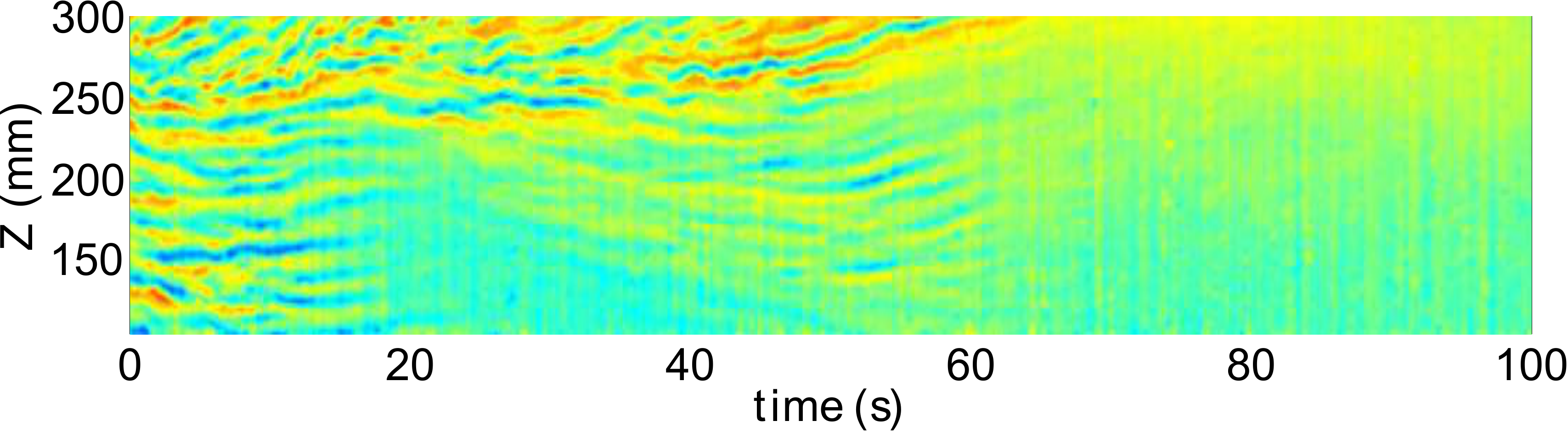}(b)
   \includegraphics[width=0.93\textwidth]{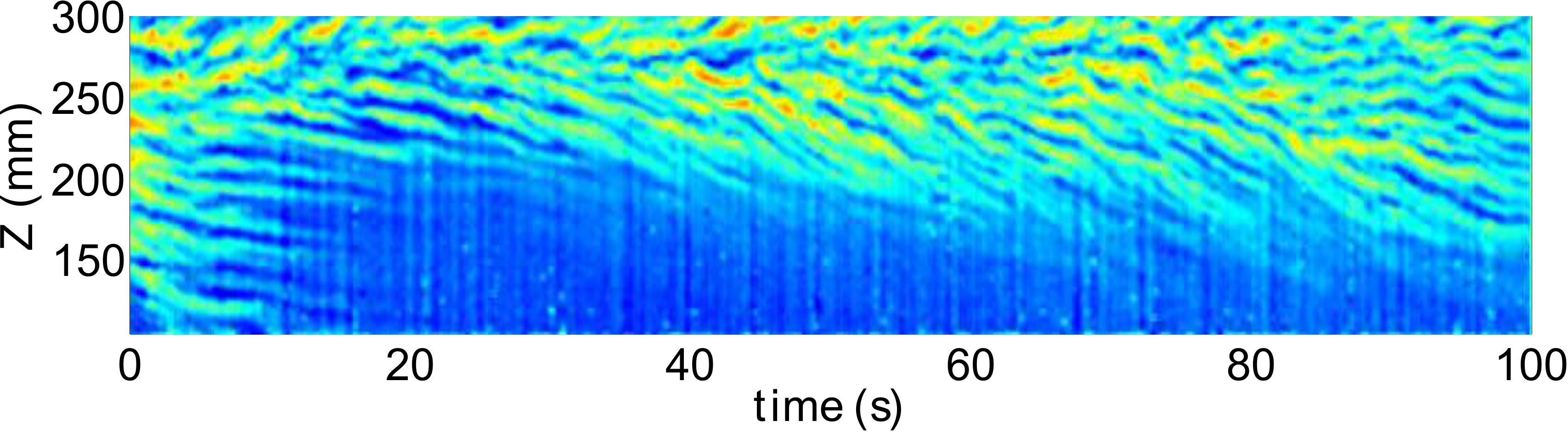}(c)
   \includegraphics[width=0.93\textwidth]{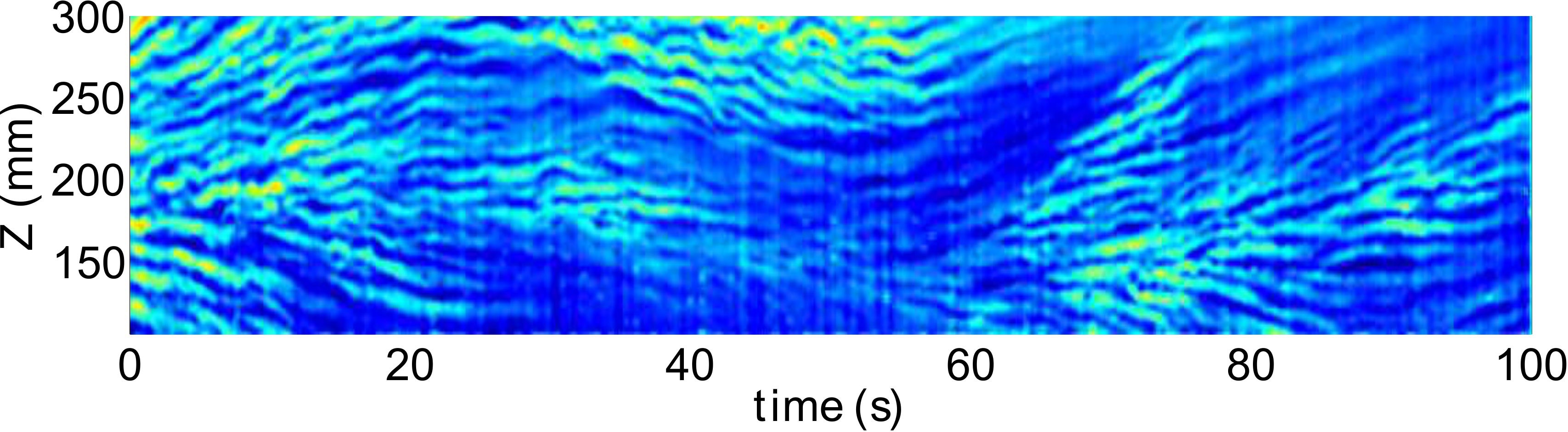}(d)
  \includegraphics[width=0.7\textwidth]{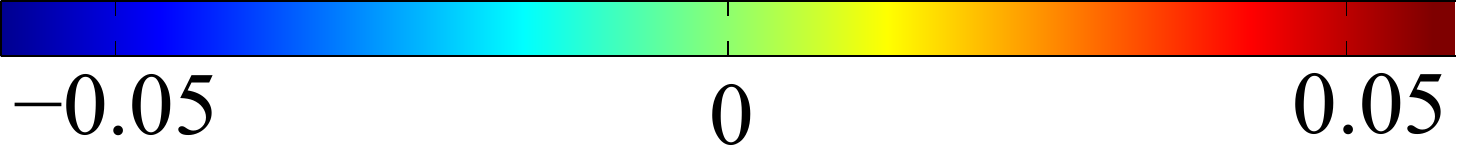}
\end{minipage}
	\caption{Left: general sketch of the plane Couette flow set-up (a). Right: $U_x$ space-time (t,z) diagrams obtained after a water jet has perturbed the flow (see text for details). $U=83.5$~mm.s$^{-1}$, $Re=330$, $2h=7.5$~mm. Measurements are performed at $y/h\sim 0$, $y/h \sim 0.4$ and $y/h \sim 0.5$ from (b) to (d). Color online.}
	\label{setup}
\end{figure}

\begin{figure}[h!]
\centering
\begin{minipage}[c]{.48\linewidth}
	\begin{center}
       \includegraphics[width=0.95\textwidth]{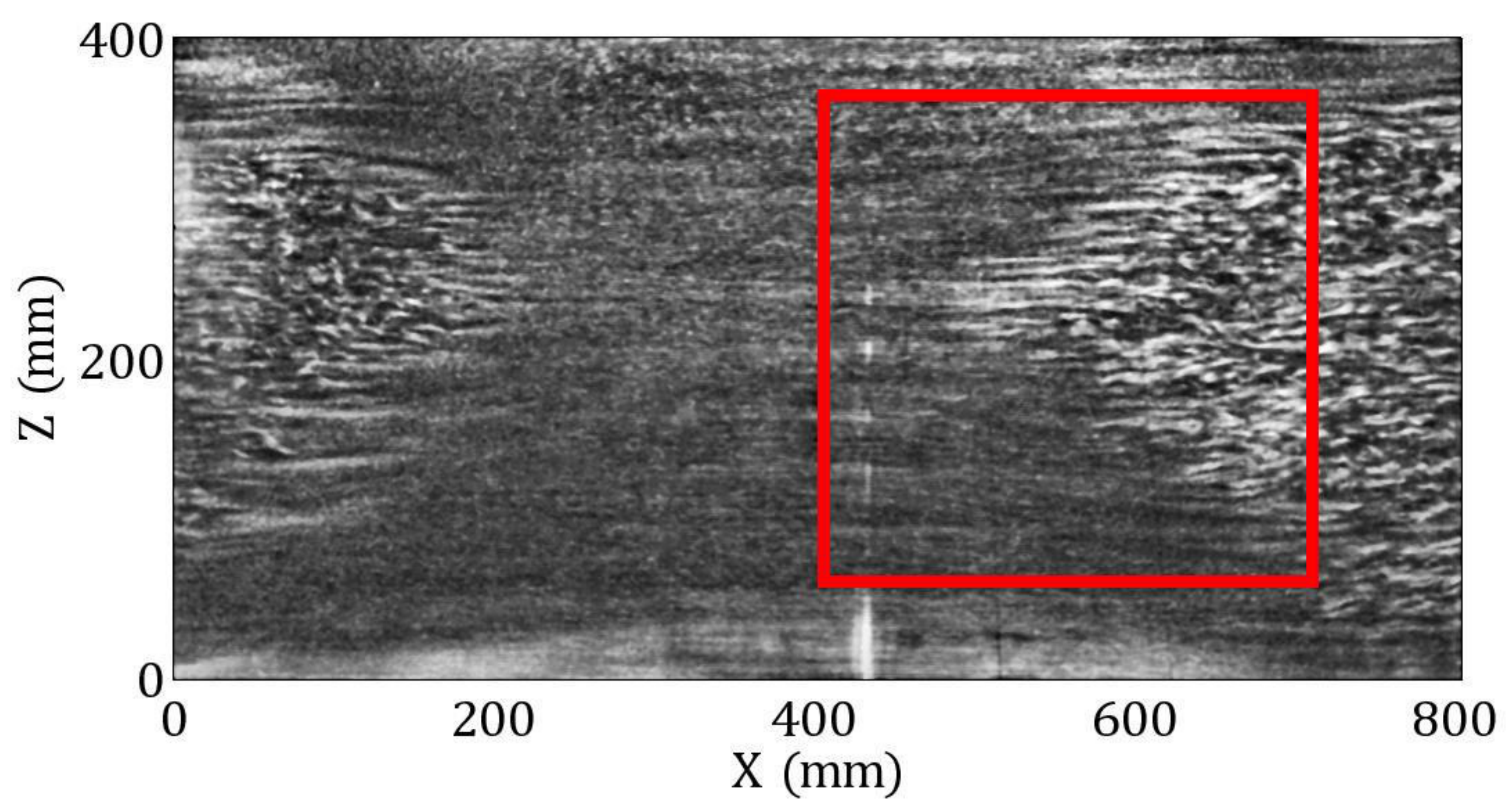}\\
       (a)
\end{center}
\end{minipage}
\begin{minipage}[c]{.48\linewidth}
	\begin{center}
\includegraphics[width=0.95\textwidth]{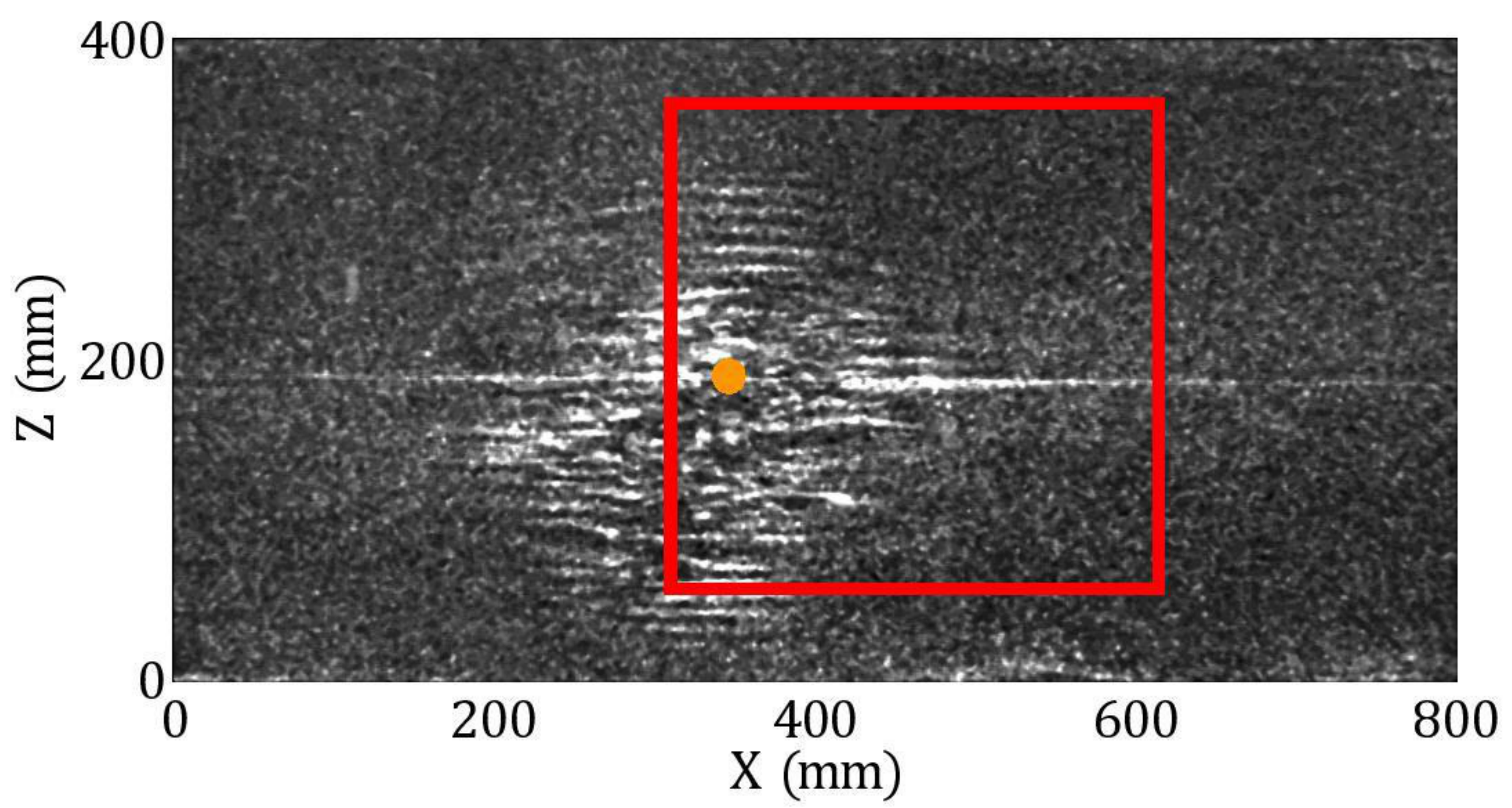}\\
(b)
\end{center}
\end{minipage}
\caption{Snapshots obtained from our visualisation system. PIV field of view is within the red frame. (a) step experiment: turbulence invades the laminar domain from the left and right edges (along $x$ direction) with a typical shape of an arrowhead. (b) bead experiment: a growing diamond patch of turbulence invades the laminar domain. The bead is situated at the orange dot. Visualisations are obtained from imaging an observation window of $800\times 400$~mm$^2$ fitting the area where the plane Couette profile is achieved. Flow is seeded with micron-size platelets (Iriodin) and illuminated in a $y$ plane with a laser sheet produced by a rotating polygonal mirror.}
\label{rampe}
\end{figure}

\paragraph*{Plane Couette flow:} the set-up sketched in fig.\ref{setup}-left was implemented to approach the ideal plane Couette flow developing between two infinite parallel plates. An endless $0.25$~mm thick plastic belt links two main cylinders. One of them is connected to a brush-less servo-motor which drives the system through a gear-reducer of ratio $9$. This motor can achieve a maximal speed of $2500$~rpm. Speed and acceleration are controlled with an accuracy of $2/1000$. Four smaller cylinders guide the belt and enable to continuously adjust the gap between the two walls moving in opposite directions. This set-up is immersed in a tank filled with water. Water temperature is monitored by a thermo-couple with a $0.5  \ensuremath{^\circ} $ accuracy. The Reynolds number is defined as $Re=Uh/\nu$, where $\nu$ is the kinematic viscosity of the fluid, $U$ the belt velocity and $h$ the half-gap between the two faces of the belt. Work presented here corresponds to a gap of $2h=7.5$~mm. Due to the experimental geometry, plane Couette profile is achieved within a $L_x=800$~mm (along $x$) $\times$ $L_z=400$~mm (along $z$) area fitting our visualisation field of view. The corresponding aspect ratios are $\Gamma_x=L_x/h=107$ and $\Gamma_z=L_z/h=53$. $x$, $y$ and $z$ and $U_x$, $U_y$ and $U_z$ are respectively the streamwise, wall-normal and spanwise directions and  the associated velocities. The plane $y=0$ corresponds to the half distance between the two plates. Unless otherwise specified, the tank is closed at the top by a transparent Plexiglas lid wherever the plane Couette profile is achieved in order to enable top and bottom symmetrical boundary conditions \cite{notecite}. From various experiments not detailed here, critical Reynolds numbers have been obtained as $R_g\simeq 305$ and $R_t\simeq390$, consistently with previous experiments and well-resolved direct numerical simulations.\\
\paragraph*{Velocity measurements:} velocity measurements have been achieved by Particle Image Velocimetry (PIV). The Dantec system consists of a dual pulse Laser (Nd:YAG, $2\times135$~mJ, $4$ ns, $532$~nm) and a CCD camera (FlowSenseEO, $4$ Mpx). Image pairs are acquired at a rate of $5~$Hz. The PIV Laser is mounted above the test section on a linear traverse which allows us to measure velocities over different $y$ planes with a spatial resolution of about $1.3$~mm along $y$. The laser sheet produced from a cylindrical lens gives us $2$D velocity fields ($U_x$,$U_z$) in the streamwise/spanwise planes. The flow is seeded with particles coated with melamine and diameter ($d_p=1-20~\mu m$) that diffuses light at a shifted wavelength. A filter mounted on the camera lens is used to keep only wavelengths around this shifted wavelength, getting rid of unwanted light reflections on the vessel and plastic belt. The $1948\times 2048$ pixels$^2$ observation window corresponds to physical sizes of $196\times208$~mm$^2$ or $305\times 310$~mm$^2$. An adaptive cross-correlation processing is applied to an initial interrogation area of $64\times 64$ pixels$^2$  followed by a  final interrogation area of $32\times 32$ pixels$^2$ with a $50\%$ overlap. The corresponding spatial resolution is $2.45$~mm, {\it i.e.} $0.16h$. Each instantaneous velocity field is further filtered by removing values exceeding $1.2U$. Resulting outliers are replaced using local median filters.\\
\paragraph*{Experimental protocol:} our main goal is to focus on the coexistence of laminar and turbulent domains and on the associated large-scale flows. As explained in section \ref{sec:intro}, this situation occurs in the range of Reynolds numbers between $Re_g$ and $Re_t$ provided that a finite amplitude perturbation is added to the flow. Fig.~\ref{setup}-right presents spatio-temporal diagrams of $U_x$ obtained after a water jet has perturbed the flow. In this case, the top lid is removed and water is blowed for $100$~ms at $5$~bars.The jet nozzle is situated at $x=300$~mm and $z=380$~mm and is tilted by $45  \ensuremath{^\circ} $  with respect to the $x$ direction. The flow is initially perturbed at large scales, but after a $10$~s transient, the initial perturbation shrinks and a localized turbulent spot remains. From one realization to another, the fate of this spot drastically changes from decay to monotonic growth and even complex spatio-temporal dynamics. The three examples presented here illustrate the wide variety of dynamics that can be observed in the transitional range and call for more reproducible perturbations to study steady laminar-turbulent fronts. \\
To achieve such reproducible perturbations, we have used two very different protocols. The first relies on a permanent perturbation that consists of a bead positioned close to the $y/h=-0.5$ plane in the middle left of the PIV field of view (see bead position in fig.~\ref{rampe}) by means of a thin horizontal wire stretched between two vertical stalks placed between the guiding and the main cylinders. This type of permanent perturbation has already been used by Bottin and co-workers \cite{bottin97_PRL}. The wire diameter is $\phi_{wire}/h=0.1$ and the bead diameter is $\phi_{bead}/h=0.8$. As explained in \cite{bottin98_POF}, a thin wire parallel to the streamwise direction does not influence the flow. As already mentioned by Bottin and co-workers, the bead diameter does not strongly affects the qualitative results and provided $Re>Re_g$, no intermittency is observed but rather only growing spots are triggered. We studied the influence of the stalks on the transition to turbulence in our system when the bead is not present: their influence is hardly noticeable when considering critical Reynolds numbers and transition time to turbulence. When turbulence develops in a formerly laminar domain, it always occurs first around the bead.\\
In an even simpler manner, a moving laminar-turbulent interface can be obtained by what we refer to hereafter as a step: a sudden increase of $Re$ from a value below $Re_g$ to a value above it. Practically, the belt is set into motion with an acceleration of $300$~mm.s$^{-2}$ from $Re=0$ to $Re=450$, {\it i.e.} above $Re_t$ so that the final state is expected to be fully turbulent. The same storyline always occurs: turbulence appears from the side edges as arrowhead shaped fronts as illustrated in fig.~\ref{rampe}. Actually, due to the experimental geometry (see fig.~\ref{setup}), the Reynolds number is higher between the guiding and the main cylinders so that turbulence is first triggered in these regions. For this type of experiments, the PIV field of view (red rectangle in fig.~\ref{rampe}) is shifted to the right of the set-up, hence we observe a turbulent front showing up from the right and slowly overrunning all the field of view by moving from right to left.
\subsection{Post-processing}
\paragraph*{Measured quantities:} to enlighten structures present in the flow, we use the autocorrelation of $U_x$ defined as:
\begin{equation}
R_{uu}(x,\Delta z,t)=\frac{1}{R_{0}}
\int_{-\infty}^{+\infty} U_x(x,z+\Delta z,t)U_x(x,z,t)dz
\end{equation}
with $R_{0}=R_{uu}(x,0,t)$. $R_{uu}(x,\Delta z,t)$ is further averaged over $x$ and $t$ to obtain $R_{uu}(\Delta z)$.
\\
Spatial spectral analysis is performed by computing $U_x$ and $U_z$ pre-multiplied power spectra obtained from $\widehat{U_{x/z}}(k_x,k_z)$, the two-dimensional $U_x$ and $U_z$ Fourier transforms. The corresponding power spectra, $E_{x/z}(k_x,k_z)=|\widehat{U_{x/z}}(k_x,k_z)|^2$ are azimuthally averaged to obtain $E_{x/z}(k)$, with $k=2\pi/\lambda=\sqrt{k_x^2+k_z^2}$. $k_x$ and $k_z$ being the wave numbers along $x$ and $z$ respectively. Pre-multiplied power spectra are finally obtained as $kE_{x/z}(k)$.
\begin{figure}
  \begin{minipage}[c]{.24\linewidth}
  \begin{center}
  \includegraphics[clip=true,trim=30 50 90 0,width=1\textwidth]{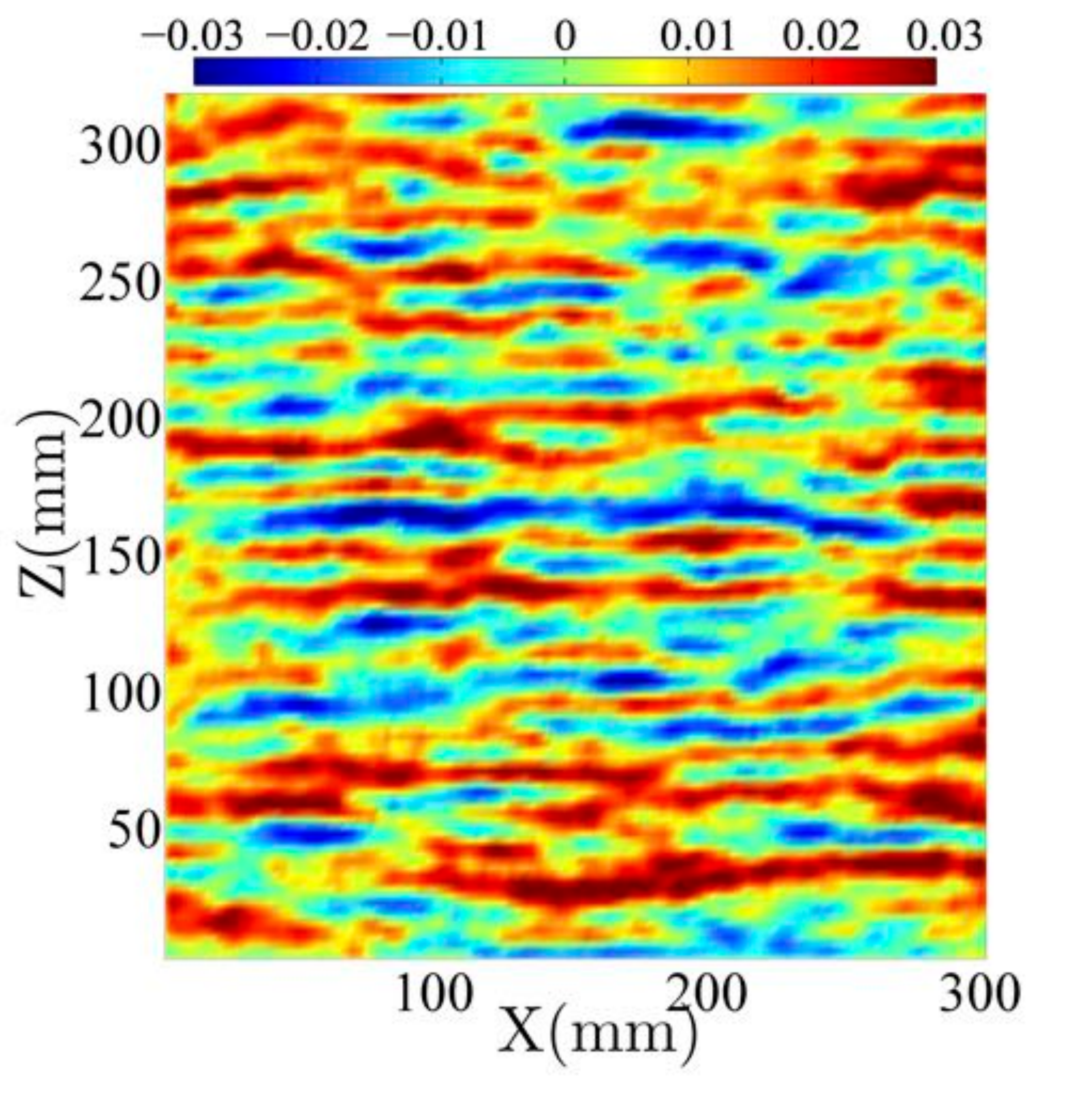}\\
  (a)
  \end{center}
  \end{minipage}
  \begin{minipage}[c]{.24\linewidth}
  \begin{center}
  \includegraphics[clip=true,trim=20 20 90 0,width=1\textwidth]{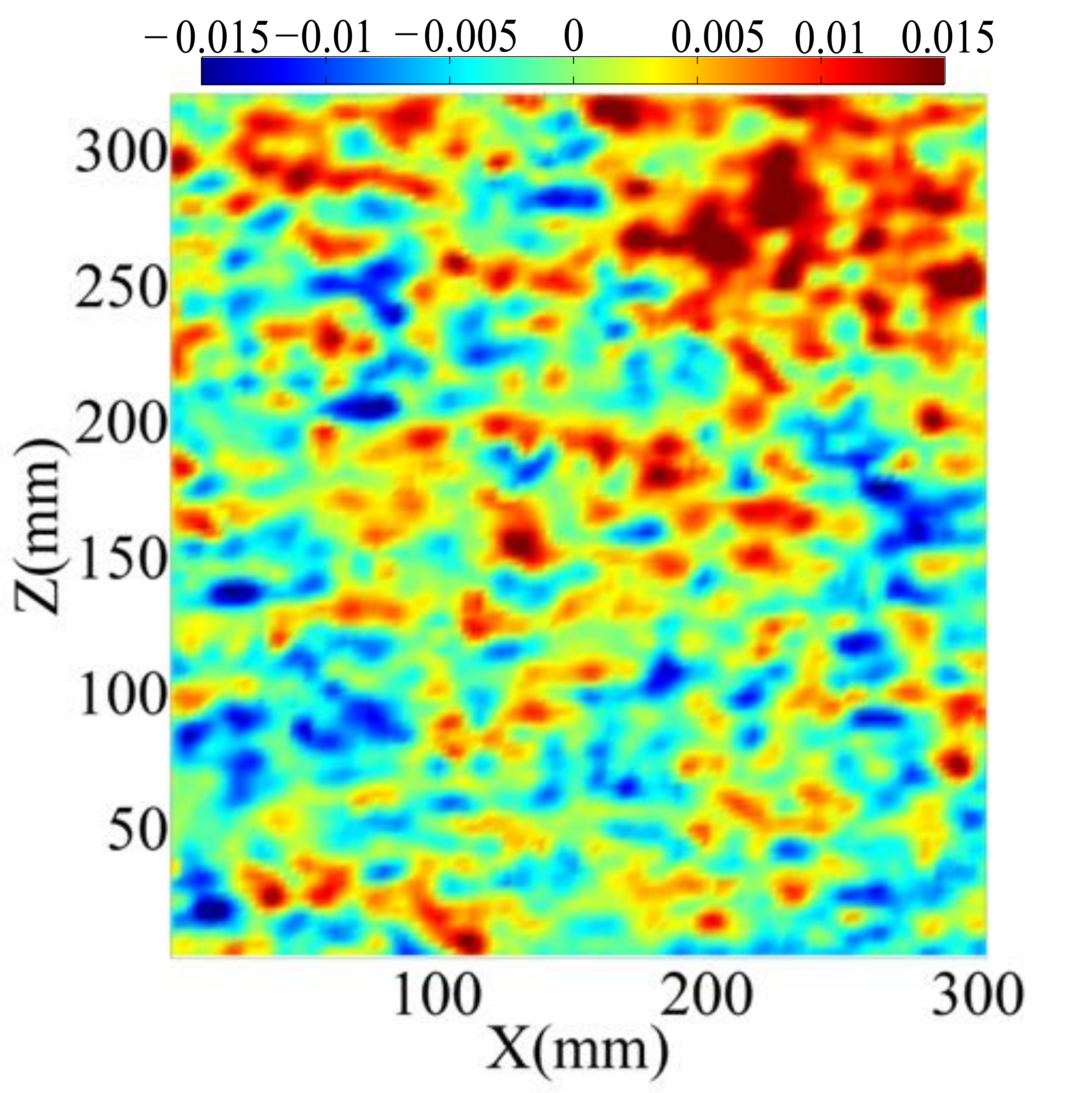}\\
  (b)
  \end{center}
  \end{minipage}
  \begin{minipage}[c]{.24\linewidth}
  \centering  
  \includegraphics[width=1\linewidth]{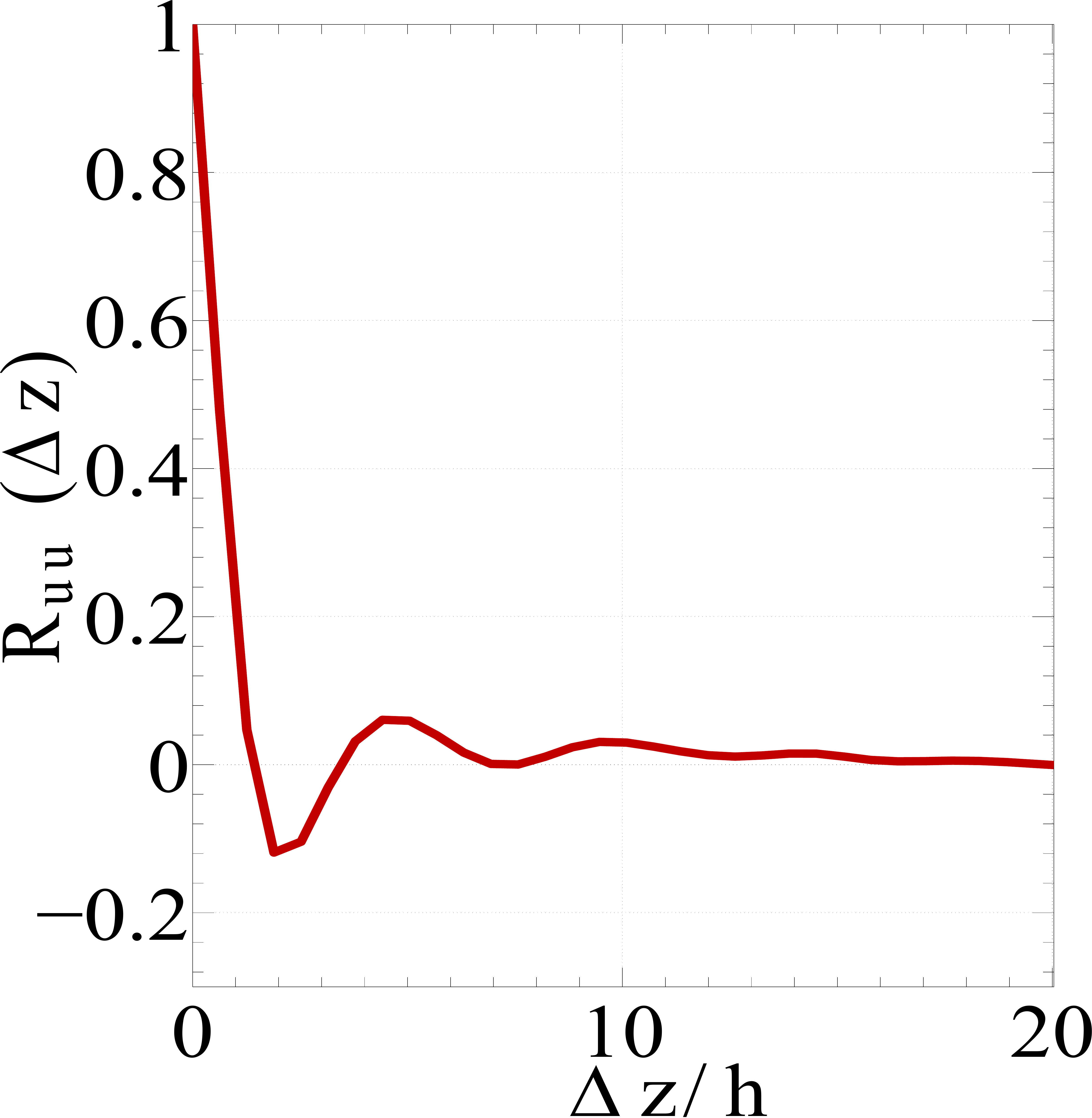}\\
  (c)
  \end{minipage}
  \begin{minipage}[c]{.24\linewidth}
  \centering
  \includegraphics[width=1\linewidth]{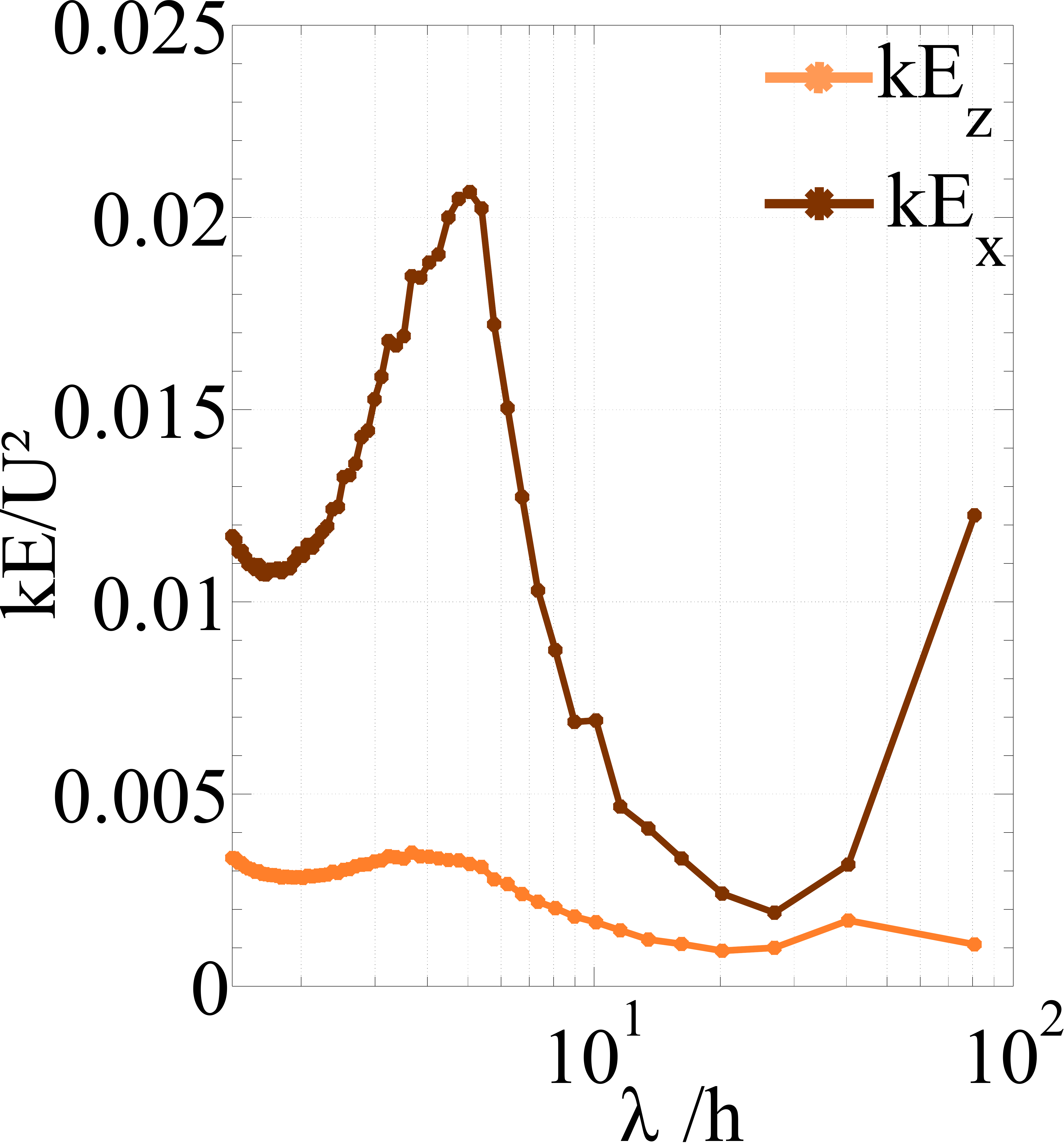}\\
  (d)
  \end{minipage}
 \caption{From left to right: velocity fields in the $x-z$ plane, $U_x$ (a) and $U_z$ (b), $U=119,4~mm.s^{−1}$, $Re=450$, $y/h\approx0$. Autocorrelation of $U_x$ along $z$ averaged along the streamwise direction and in time (c) and $1D$ time averaged power spectrum (d). Color online}
\label{turb}
\end{figure}
\paragraph*{Validation on laminar and turbulent flows:} in fig.~\ref{turb}, $U_x$ and $U_z$ fields are shown in a $(x,z)$ plane at $Re=450$ (homogeneously turbulent case). These fields are the result of a sliding average of $5$ snapshots acquired at $5$Hz. Whatever the wall-normal plane considered, the velocity fields share the same attributes. An alternation of high and low speed streaks typical of turbulent wall shear flows are seen on $U_x$. Fluctuations are also observed on $U_z$, but their amplitude is twice lower and no obvious structure appears. $R_{uu}$ plotted in fig.~\ref{turb} (c) exhibits a tip around $2h$ giving the characteristic size of the streaks as about four times the half-gap. This typical size is in line with the numerous previous estimations in wall-bounded shear flows. The time averaged pre-multiplied spectrum $kE_x$ drawn in fig.\ref{turb} (d) has a clear peak which corresponds to streaks energy. The associated wave number is coherent with the one found from the autocorrelation. The laminar flow spectrum is much lower and does not show any peak. $R_{uu}$ for the same laminar flow (not shown) is that of a white noise giving the limitations of our PIV measurements.
\section{\label{sec:res}Results}
\subsection{Evidence of large scale flows}\label{sec:LSF}
\begin{figure}[!b]
\centering
 \begin{minipage}[c]{.48\linewidth}
 \begin{center}
 \includegraphics[width=0.8\textwidth]{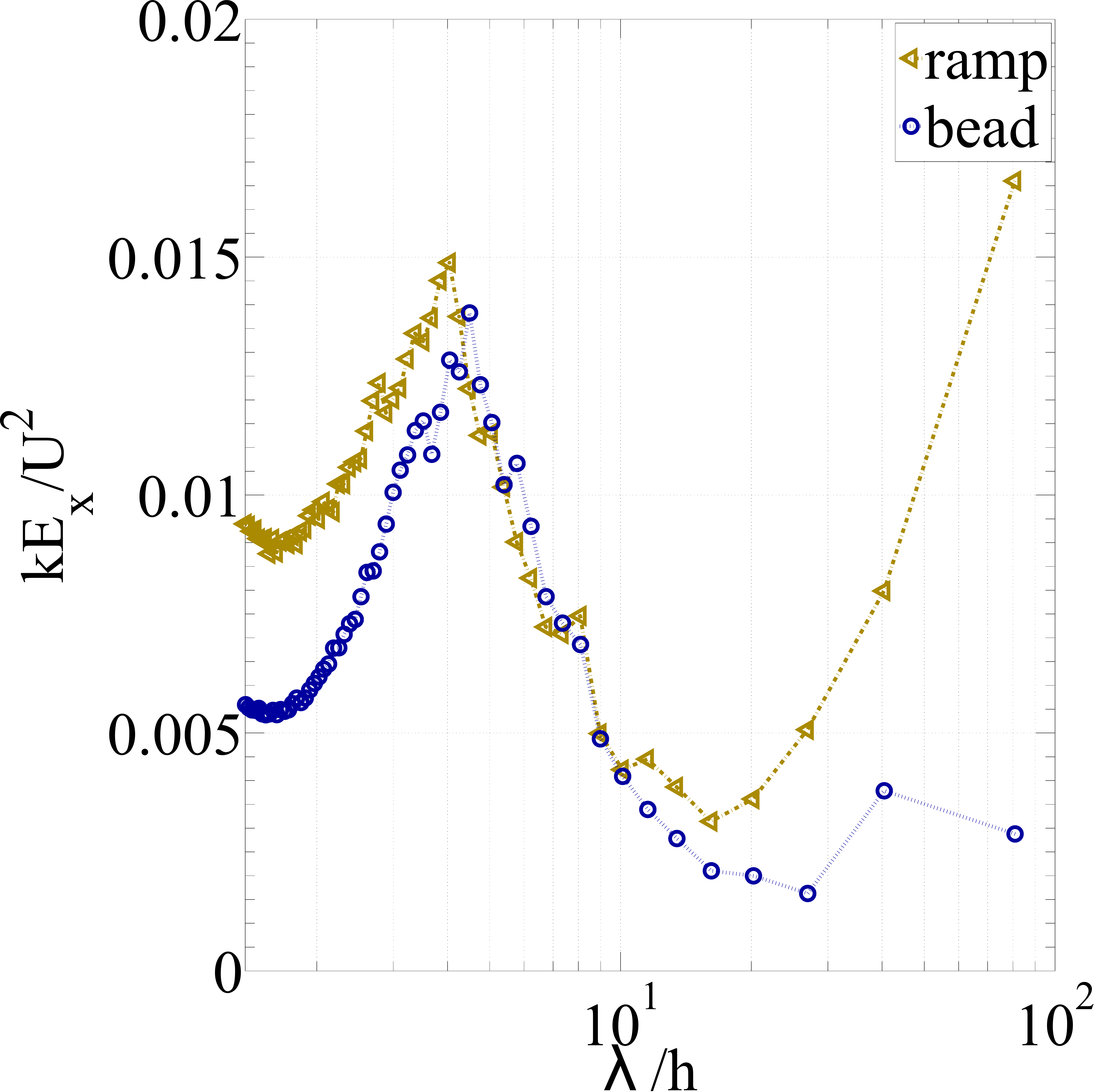}\\þ
 (a)
 \end{center}
 \end{minipage}
 \begin{minipage}[c]{.48\linewidth}
 \begin{center}
 \includegraphics[width=0.8\textwidth]{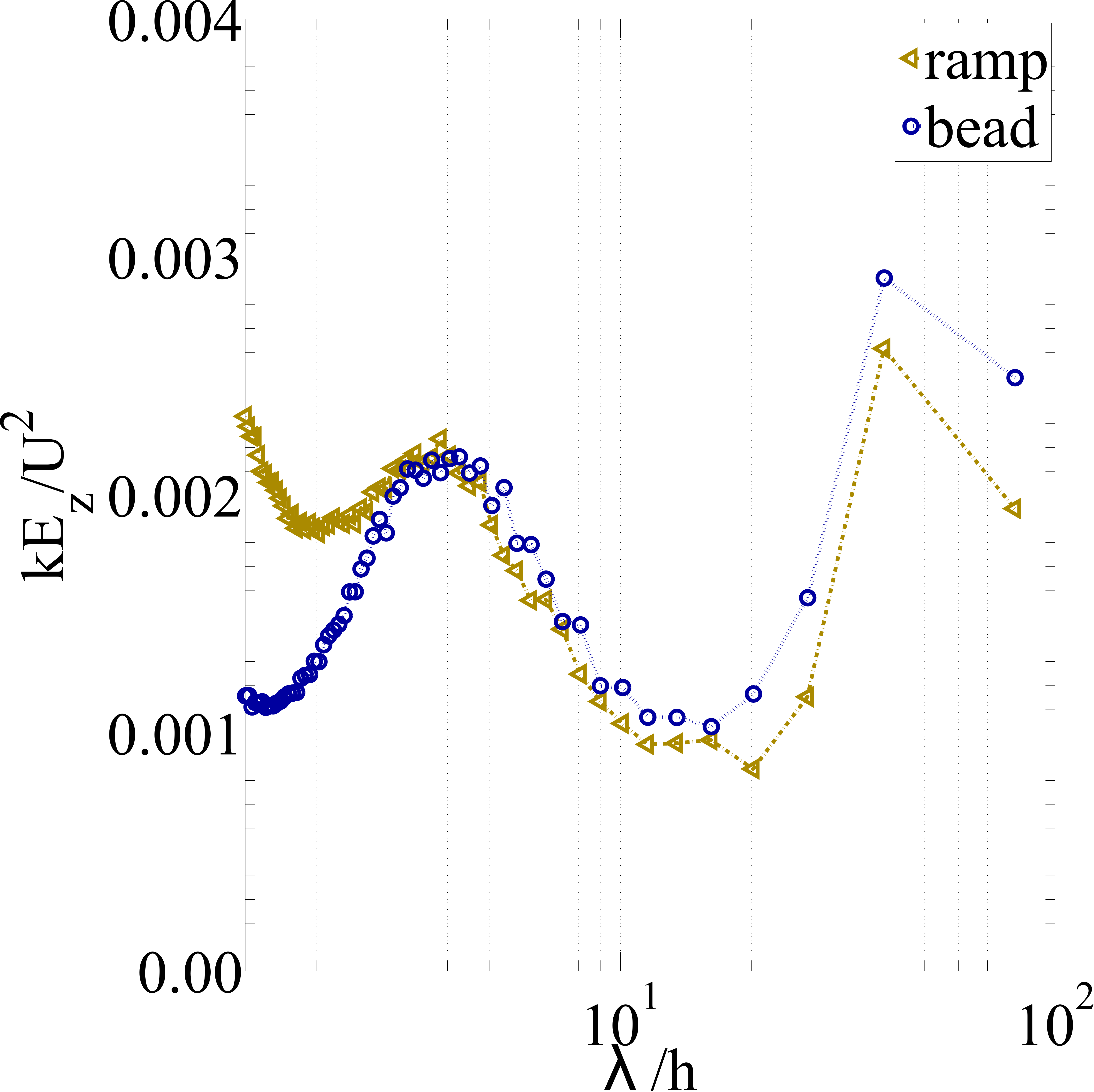}\\
 (b)
 \end{center}
 \end{minipage}
 \includegraphics[width=0.45\textwidth]{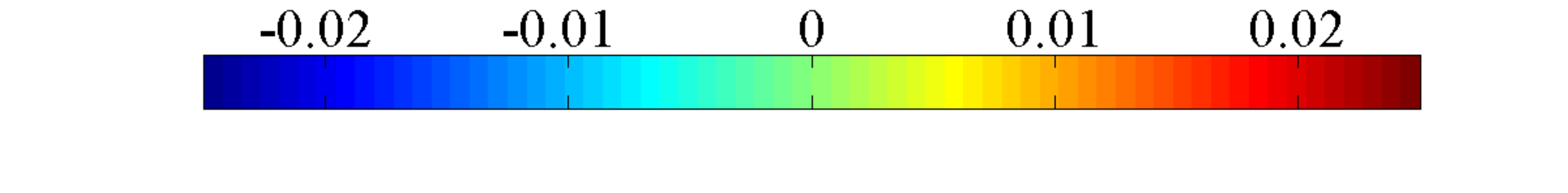}
 \includegraphics[width=0.45\textwidth]{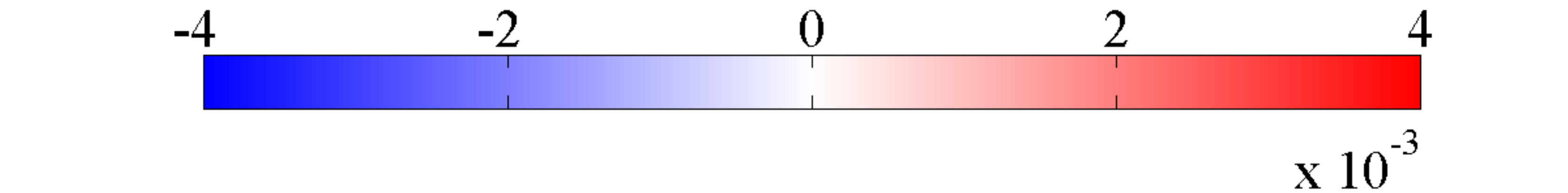}
 \begin{minipage}[c]{.24\linewidth}
 \begin{center}
 \includegraphics[clip=true,trim=30 50 90 100,width=1\textwidth]{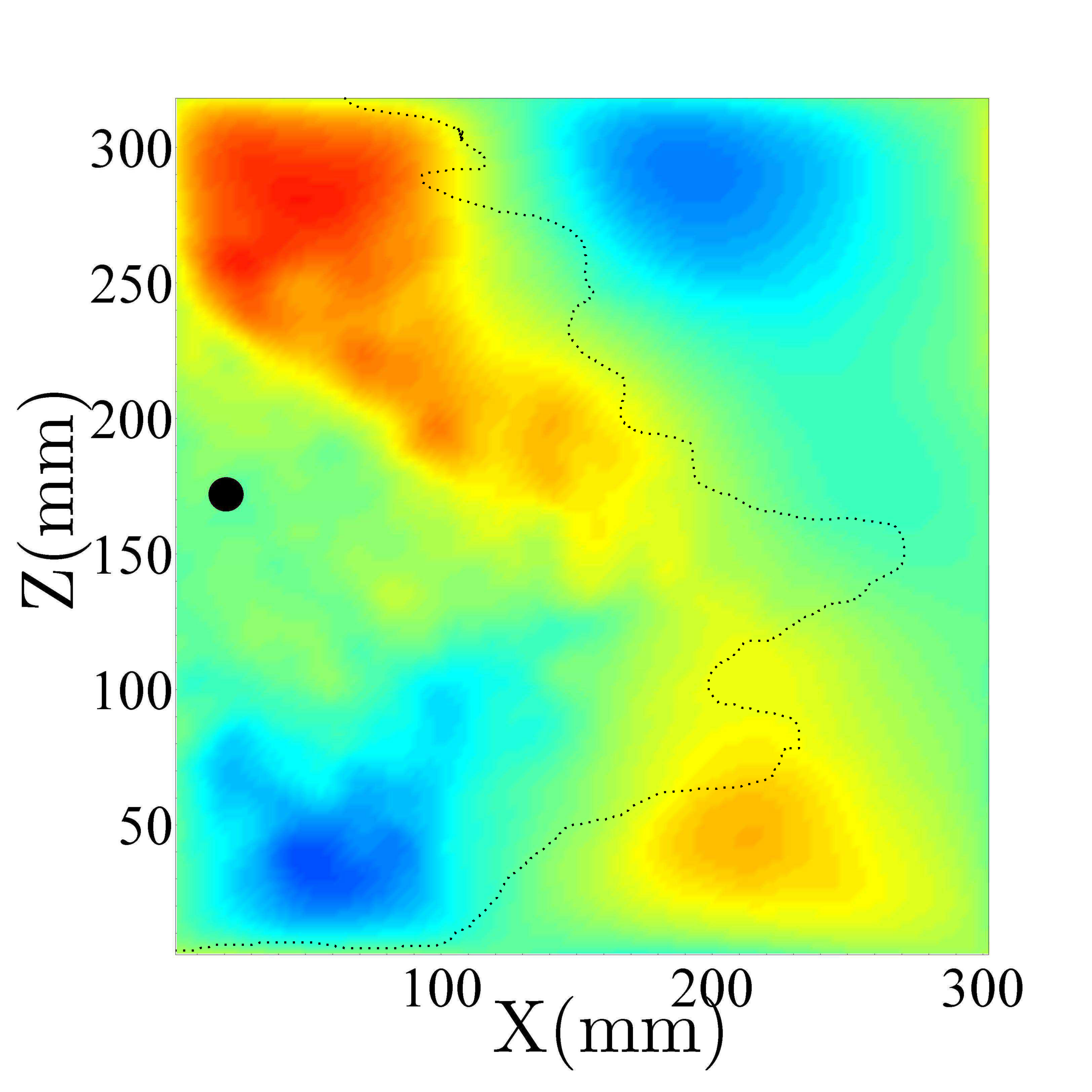}
 \\
 (c)
 \end{center}
 \end{minipage}
 \begin{minipage}[c]{.24\linewidth}
 \begin{center}
 \includegraphics[clip=true,trim=30 50 90 100,width=1\textwidth] {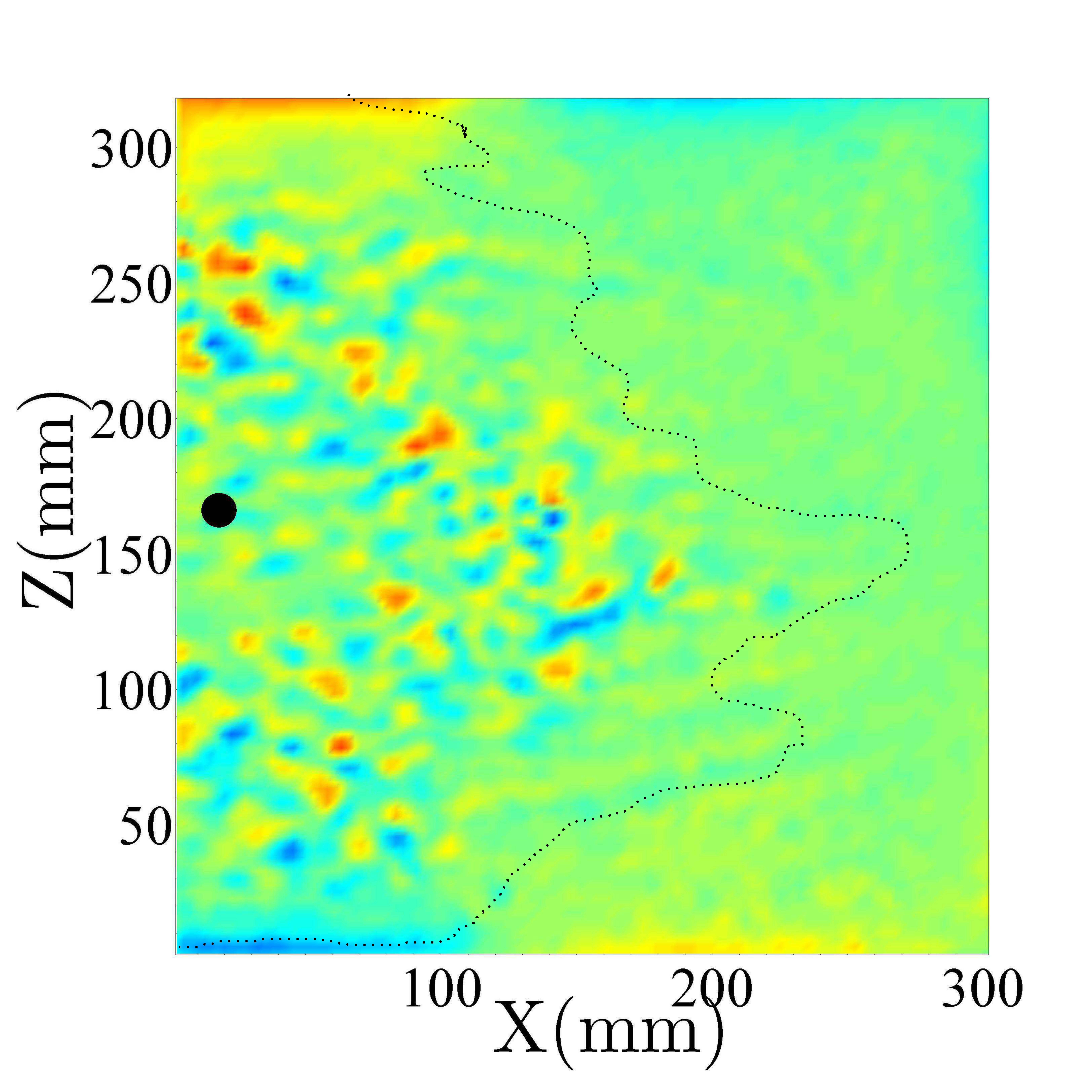}
 (d)
 \end{center}
 \end{minipage}
 \begin{minipage}[c]{.24\linewidth}
 \begin{center}
 \includegraphics[clip=true,trim=30 50 90 100,width=1\textwidth]{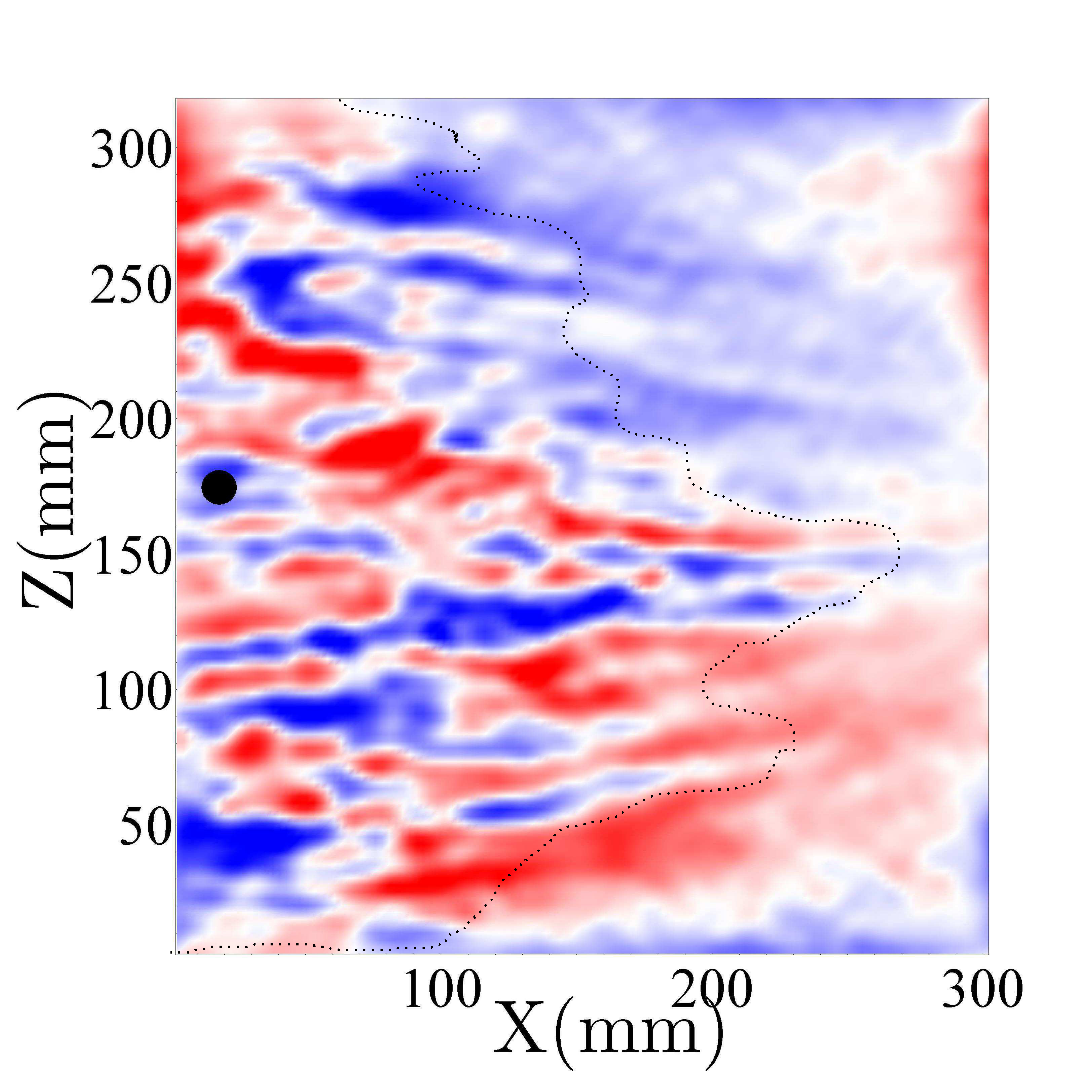}
 \\
 (e)
 \end{center}
 \end{minipage}
 \begin{minipage}[c]{.24\linewidth}
 \begin{center}
 \includegraphics[clip=true,trim=30 50 90 100,width=1\textwidth]{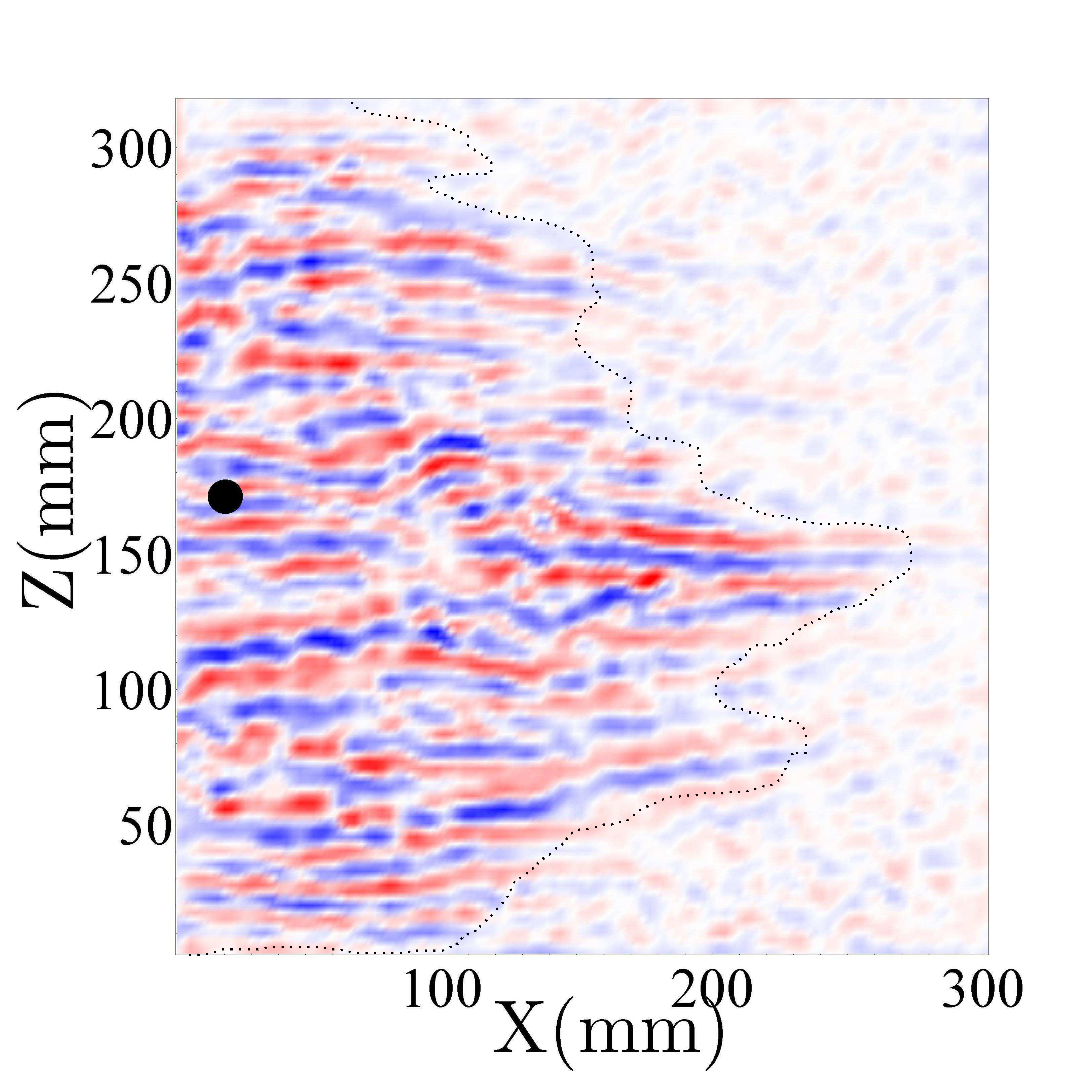}
 \\
 (f)
 \end{center}
 \end{minipage}
\caption{Top: dimensionless $U_x$ (a) and $U_z$ (b) premultiplied power spectra for typical step (triangles) and bead (circle) experiments. Measurement are done at $y/h \sim 0.2$ and $Re=450$ for the step and at $y/h\sim 0$ and $Re=403$ for the bead. Bottom: velocity and vorticity field in $(x,z)$ plane for a bead experiment. $U_z$ at large (c) and small (d) scale, $10\omega_y$ at large (e) and $\omega_y$ at small (f) scale. Measurements are done at $y/h\sim 0$, $Re=347$. The bead is situated at the black dot. A dashed line corresponding to the turbulent spot contour obtained from (f) is shown for all snapshots. Color online.}
\label{fig:scalesep}
\end{figure}

For both perturbation protocols (steps and localized bead), the flow is fully laminar at the beginning of the experiment. Turbulence appears either at the streamwise edges of the experiment or around the bead as illustrated in fig.~\ref{rampe}. Then, it gradually invades more and more space before reaching a stationary turbulent fraction (see caption for details). This turbulent fraction may be equal to one provided the final Reynolds number is higher than $Re_t$ or less than one otherwise. In any case, the flow is most of the time in a state where turbulent and laminar areas coexist. Fig.~\ref{fig:scalesep} presents premultiplied power spectra associated to $U_x$ (a) and $U_z$ (b) obtained from both types of experiments. The spectra are calculated at times corresponding to intermediate turbulent fractions and are averaged over $2$~s ($10$ consecutive samples). Two peaks are visible, one around $\lambda/h\simeq4-5$ and the other around $\lambda/h\simeq40$. The former corresponds to the streaks as discussed in section \ref{sec:meth}, the latter to large-scale structures. The peak associated with streaks is five times more intense on the streamwise component than on the spanwise one. This ratio is consistent with the case where the flow is homogeneously turbulent as discussed in \ref{sec:meth}. Regarding the large-scale wavelength, the same energy level is observed for both components.\\
We have performed our measurements in various $y/h$ planes. Regarding the streaks energy, the peak position and intensity are unaffected by the measurement plane while the energy distribution around the large wave number is. Large-scale peak associated to $kE_x$ is more pronounced in planes closer to the belt. $kE_z$ spectra are almost unaffected by the measurement position. To summarize, it is easier to detect large-scale flows on $kE_z$ than on $kE_x$, and for $kE_x$, large-scale flows have a stronger signature on measurements performed on planes shifted from the midplane.\\
The scale separation allows us to define without ambiguity a cut-off wavelength at $\lambda_c/h=24$ to extract the large-scale and small-scale flows with $4^{th}$ order Butterworth low-pass and high-pass filters. Fig.~\ref{fig:scalesep} presents snapshots of $U_z$ and of the wall-normal vorticity ($\omega_y$) associated with large and small scales. All snapshots correspond to the same given time along the growth of a spot around the bead at $Re=347$ when the spot fills about half the PIV field of view. The low-pass filtered vorticity and velocity fields evidence a recirculation area all around the spot which corresponds to the quadrupolar flow obtained numerically by Duguet and Schlatter\cite{duguet13_PRL} and Lagha and Manneville\cite{lagha07_POF}. The small-scale flow consists of long coherent streamwise velocity streaks visible from the vorticity (f). These structures are much less coherent along $x$ when considering $U_z$ (d). These disorganized and spatially fluctuating structures are the footprint of turbulence inside the spot and thus allow to discriminate turbulent from quasi-laminar region. Considering fig\ref{fig:scalesep}-c and \ref{fig:scalesep}-f together shows that the large-scale recirculation is present not only within the laminar area but also in the turbulent one.\\
We have shown that along the spreading of turbulence into formerly laminar plane Couette flow large and small-scale flows coexist. Small-scale features are associated with streaks within the turbulent areas while the large-scale recirculation spreads over laminar and turbulent areas. This recirculation is linked to the laminar-turbulent coexistence and is involved in the spreading of turbulence as discussed in section \ref{sec:disc}. The temporal evolution of these structures is examined in the following sections.
\subsection{Growth dynamics}\label{sec:LSFgrowth}
\begin{figure}[!b]
 \begin{minipage}[c]{.24\linewidth}
 \begin{center}
 \includegraphics[clip=true,trim=30 0 90 100,width=0.98\textwidth]{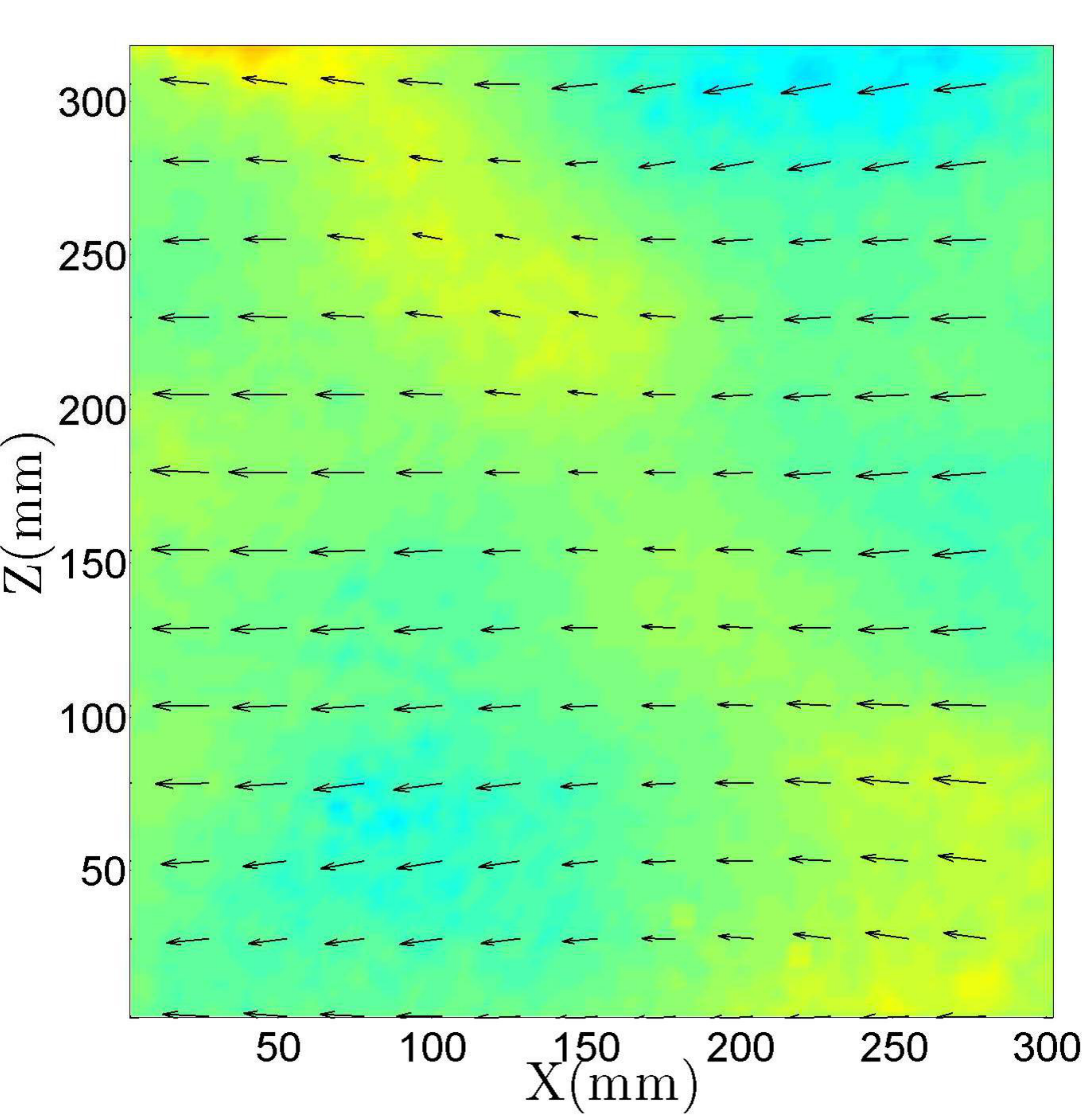}\\
 (a)
 \end{center}
 \end{minipage}
 \begin{minipage}[c]{.24\linewidth}
 \begin{center}
 \includegraphics[clip=true,trim=30 0 90 100,width=0.98\textwidth]{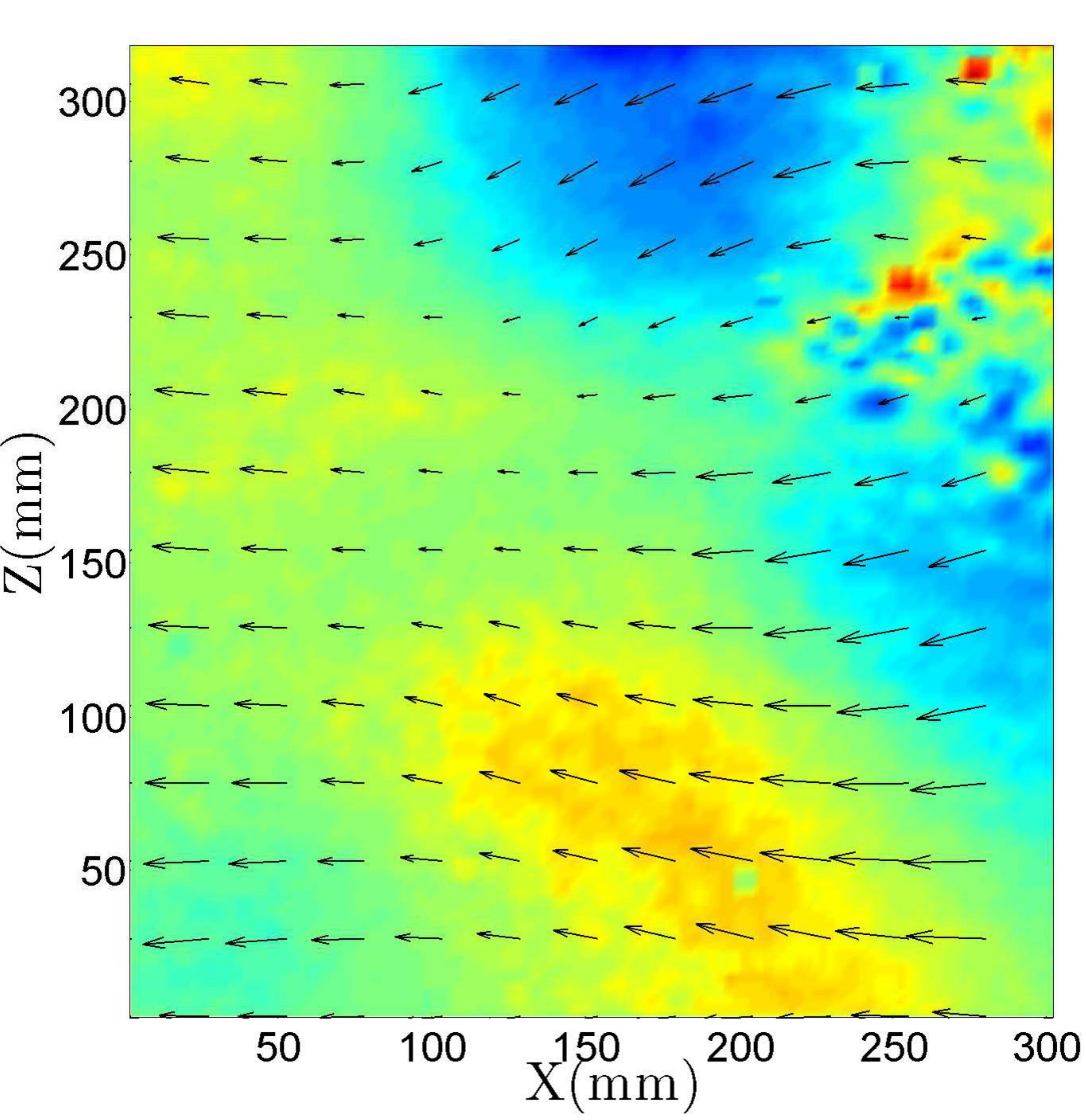}\\
 (b)
 \end{center}
 \end{minipage}
 \begin{minipage}[c]{.24\linewidth}
 \begin{center}
 \includegraphics[clip=true,trim=30 0 90 100,width=0.98\textwidth]{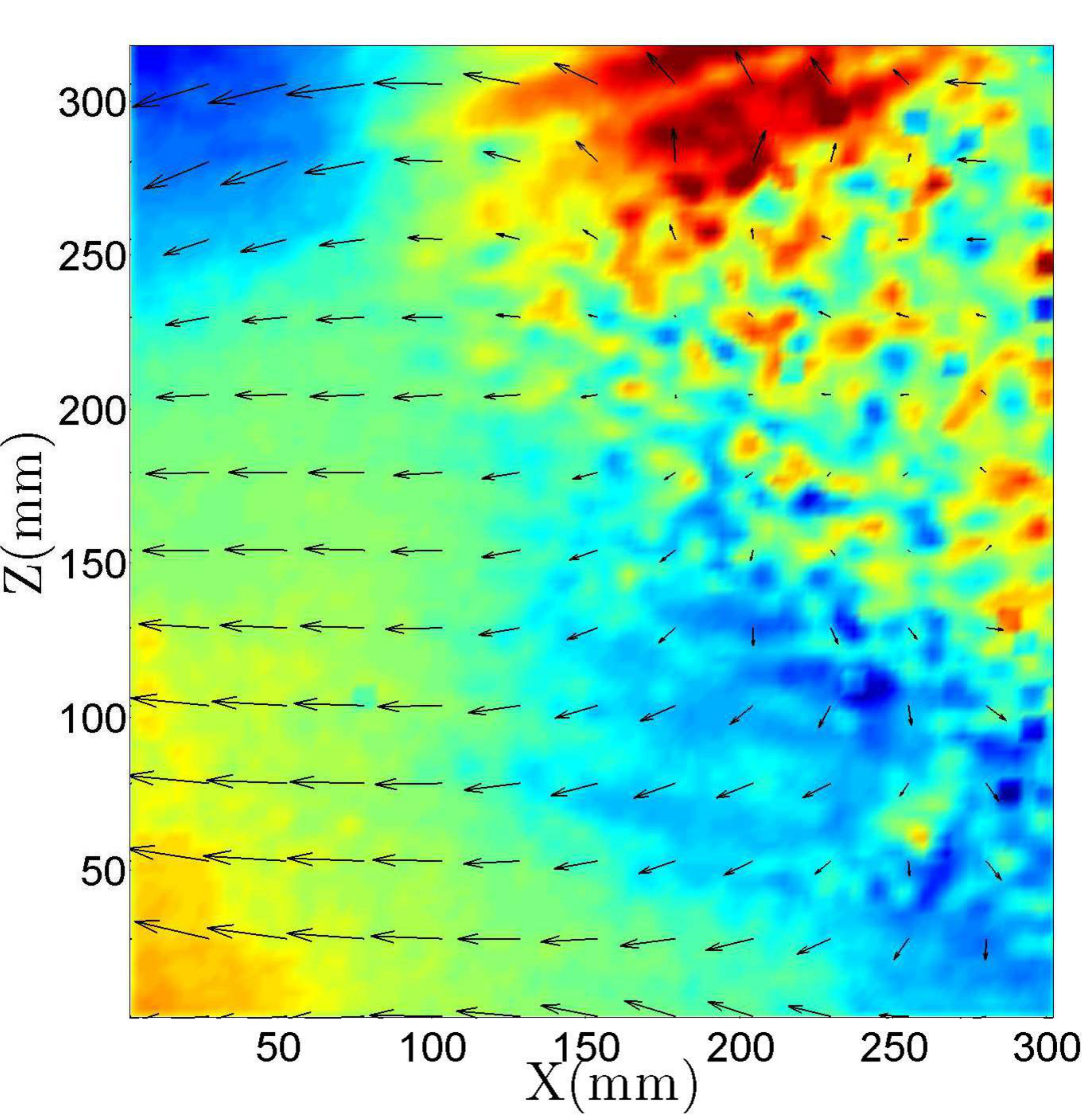}\\
 (c)
 \end{center}
 \end{minipage}
 \begin{minipage}[c]{.24\linewidth}
 \begin{center}
 \includegraphics[clip=true,trim=30 0 90 100,width=0.98\textwidth]{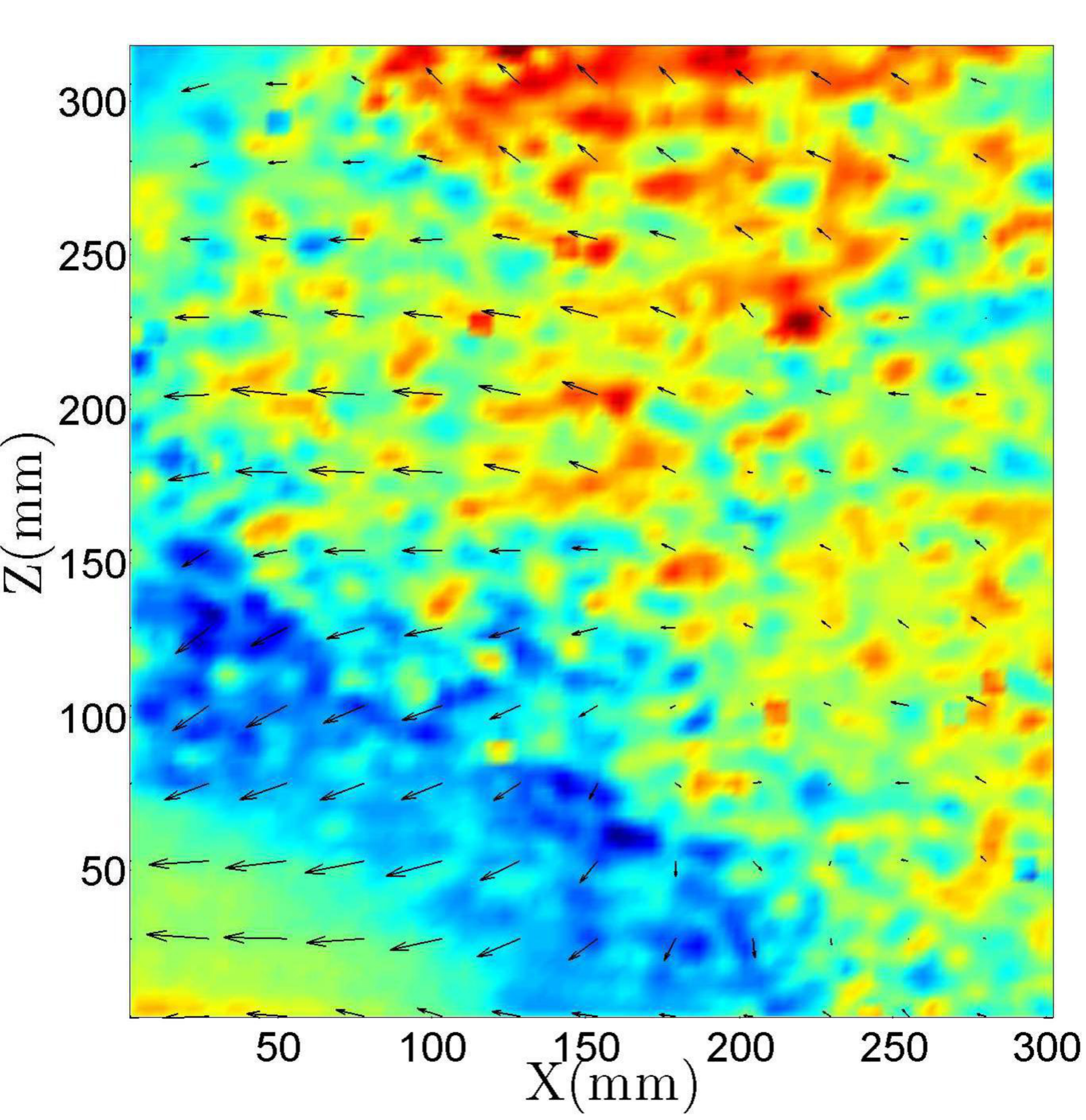}\\
 (d) 
 \end{center}
 \end{minipage}
 \begin{minipage}[c]{.5\linewidth}
 \begin{center}
 \includegraphics[width=0.75\textwidth]{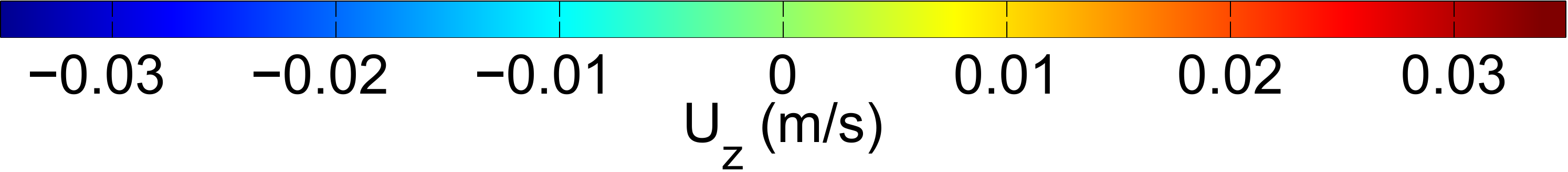}
 \end{center}
 \end{minipage}
\caption{$U_z$ snapshot during a step experiment. (a) $tU/h=62$ , (b) $tU/h=370$, (c) $tU/h=551$, (d) $tU/h=933$), Re=450. Vectors correspond to the $(U_x,U_z)$ large scale velocity field. Color online. }
\label{fig:uramp}
\end{figure}

\begin{figure}[t!]
\begin{minipage}[h]{.07\linewidth}
 \includegraphics[width=.95\textwidth]{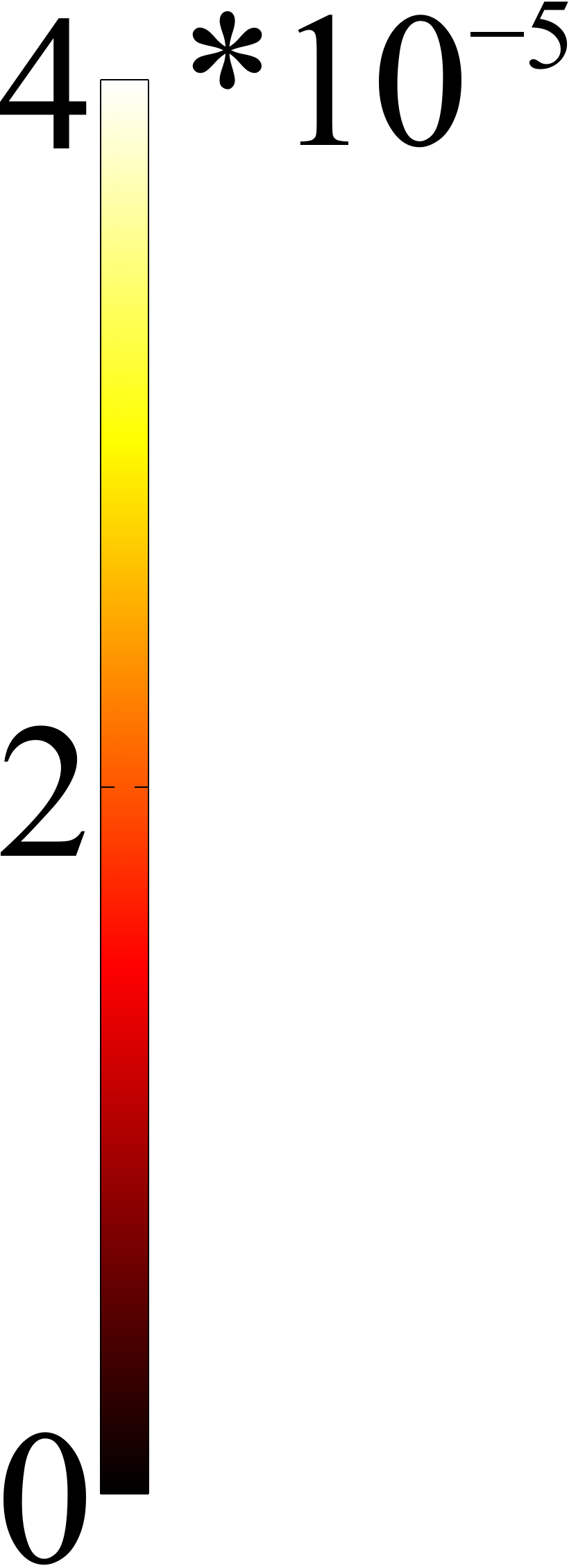}
\end{minipage}
\begin{minipage}[h]{.65\linewidth}
 \includegraphics[clip=true,trim=200 0 0 0,width=\textwidth]{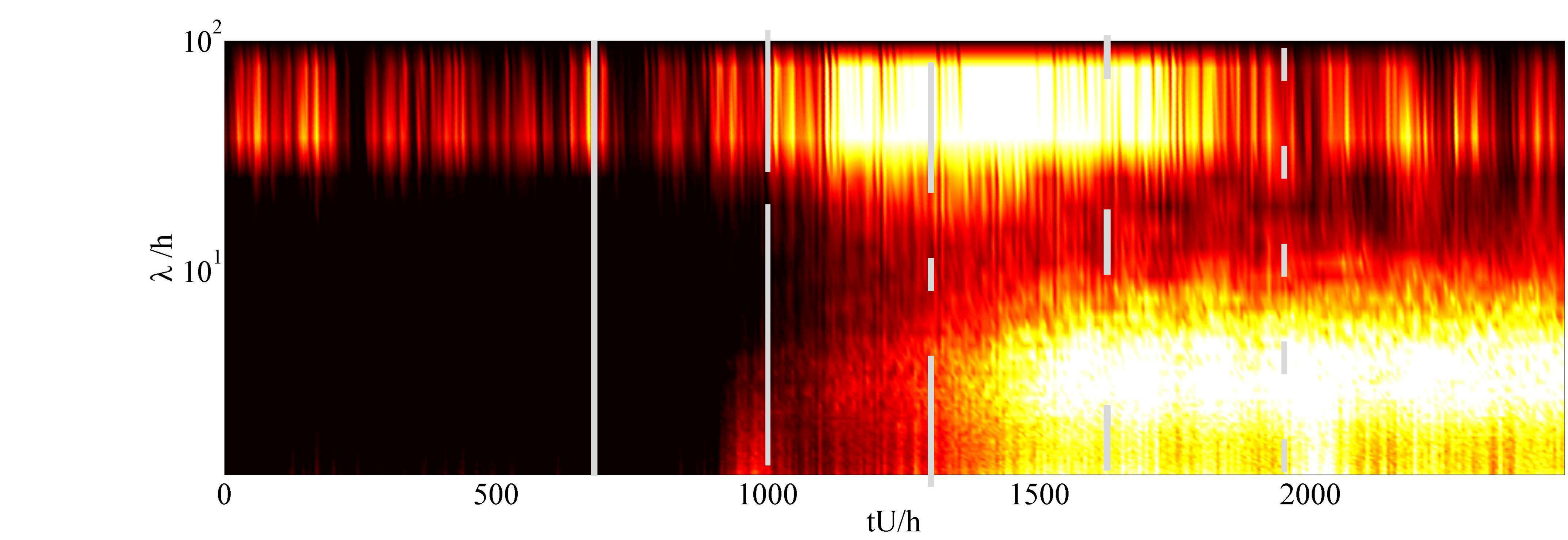}\\
 (a)
 \includegraphics[clip=true,trim=200 0 0 0,width=\textwidth]{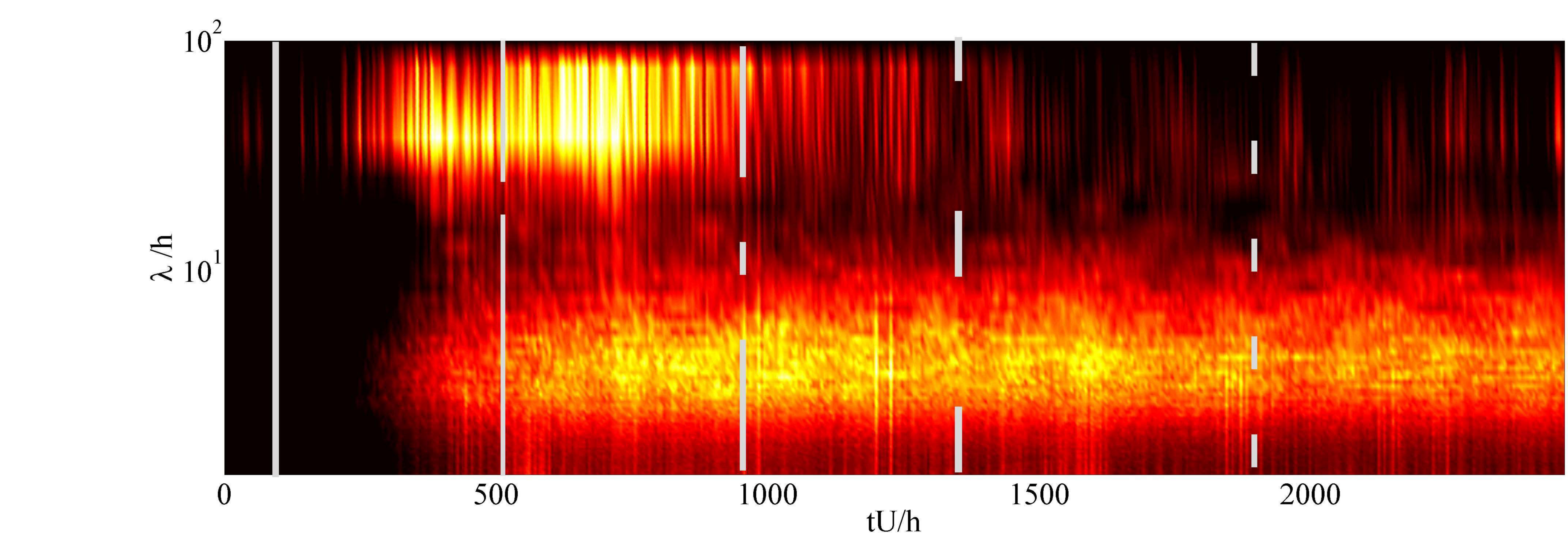}\\
 (b)
\end{minipage}
\begin{minipage}[h]{.25\linewidth}
\includegraphics[width=0.95\textwidth]{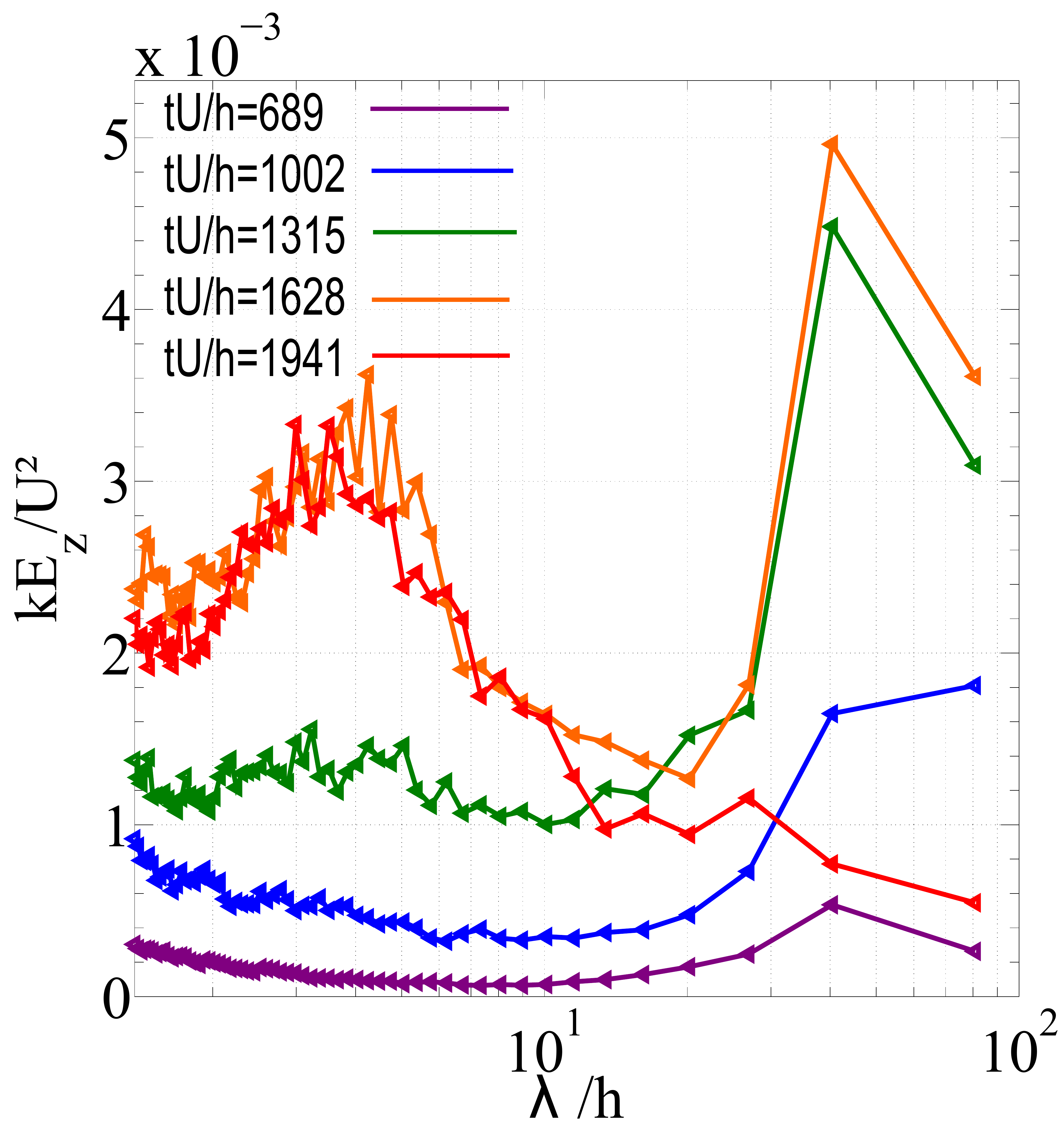}\\
(c)
\includegraphics[width=0.95\textwidth]{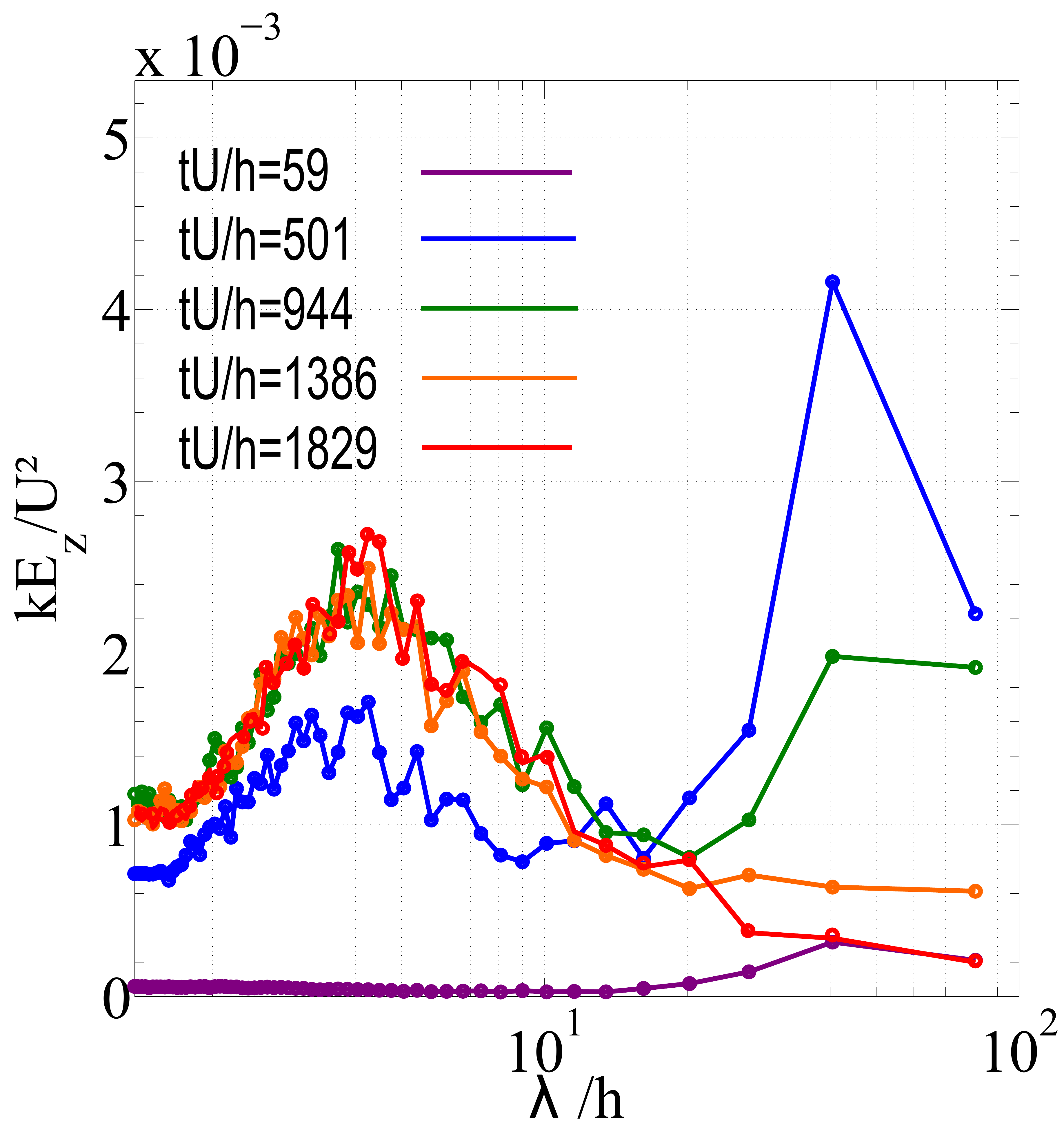}\\
(d)
\end{minipage}
\caption{Left: spectrograms of the streamwise pre-multiplied energy $kE_z$. measured at $y/h \sim 0.2$. Right: spectra extracted from the left figures at instants marked by grey lines. a-c: step at $y/h \sim 0.2$ and $Re=450$ , b-d: bead at $y/h\sim 0$ and $Re=403$. Color online}
\label{fig:Spectrogram_ez}
\end{figure}

\begin{figure}[b!]
   \begin{minipage}[c]{.32\linewidth}
		\begin{center}
  \includegraphics[width=1\linewidth]{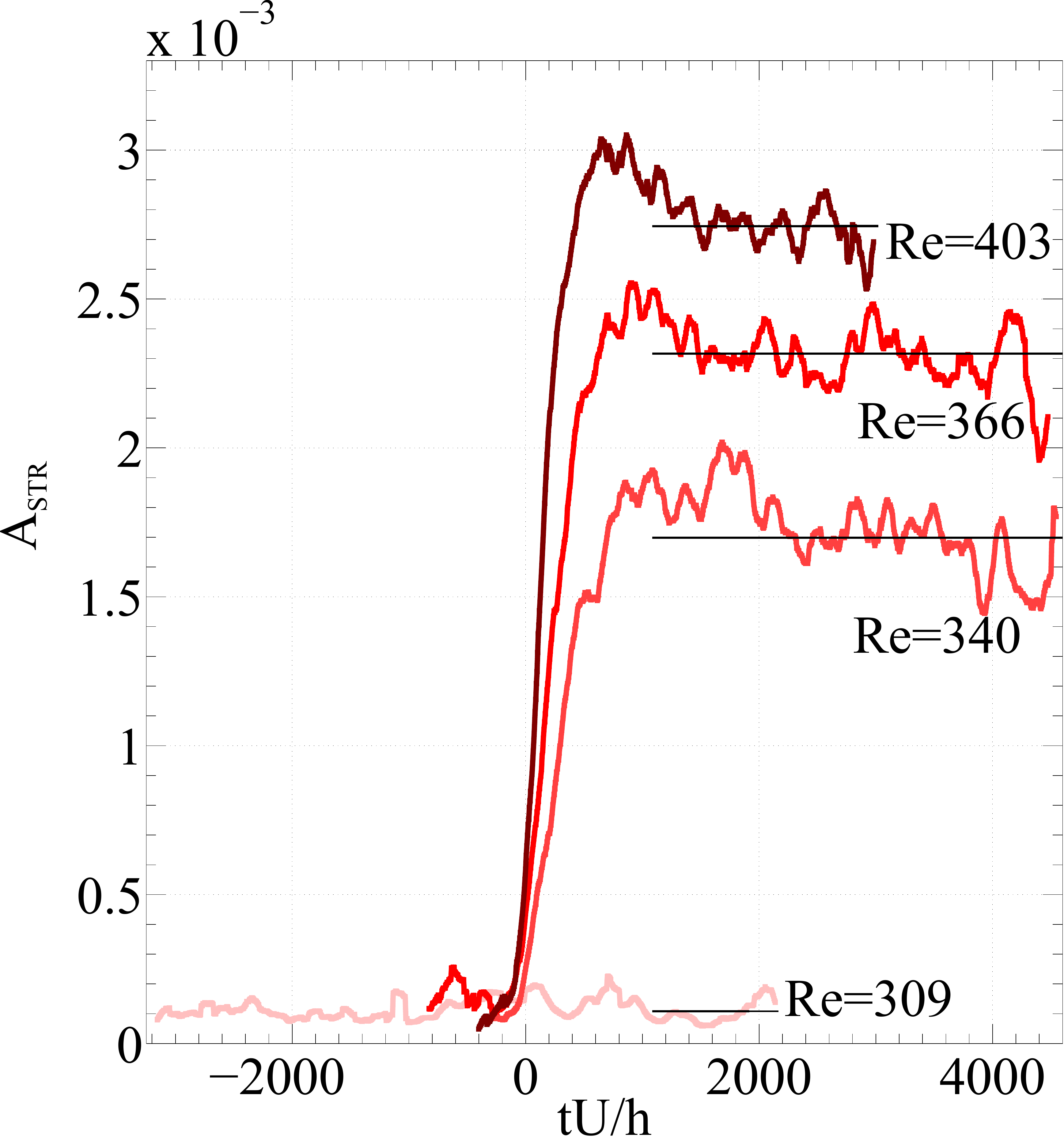}
			\\
		(a)
        \end{center}
   \end{minipage}
	   \begin{minipage}[c]{.32\linewidth}
     \centering  \includegraphics[width=1\linewidth]{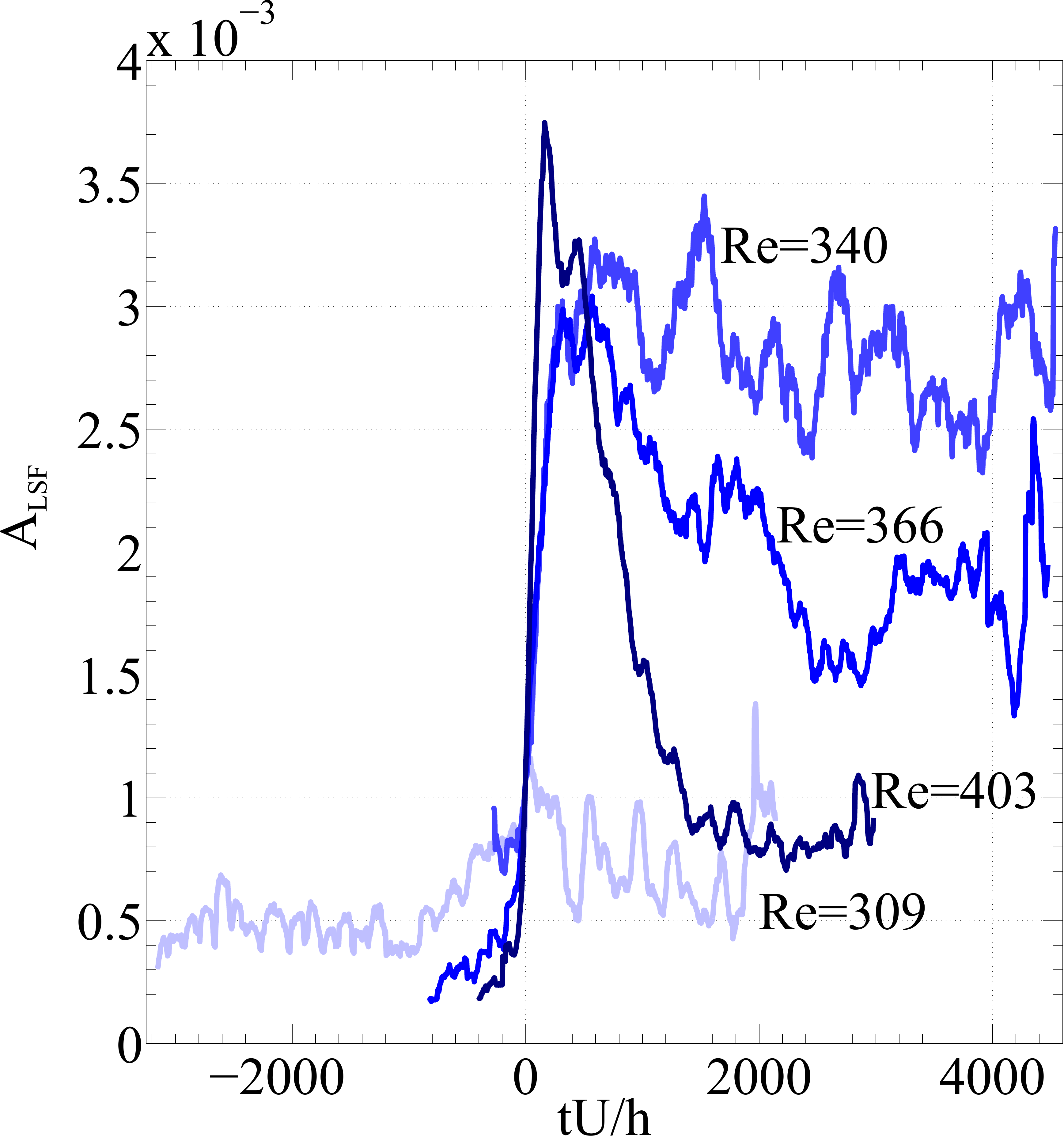}
				\\
		(b)
      \end{minipage}
				\\
   \begin{minipage}[c]{.32\linewidth}
		\begin{center}
\includegraphics[width=1\linewidth]{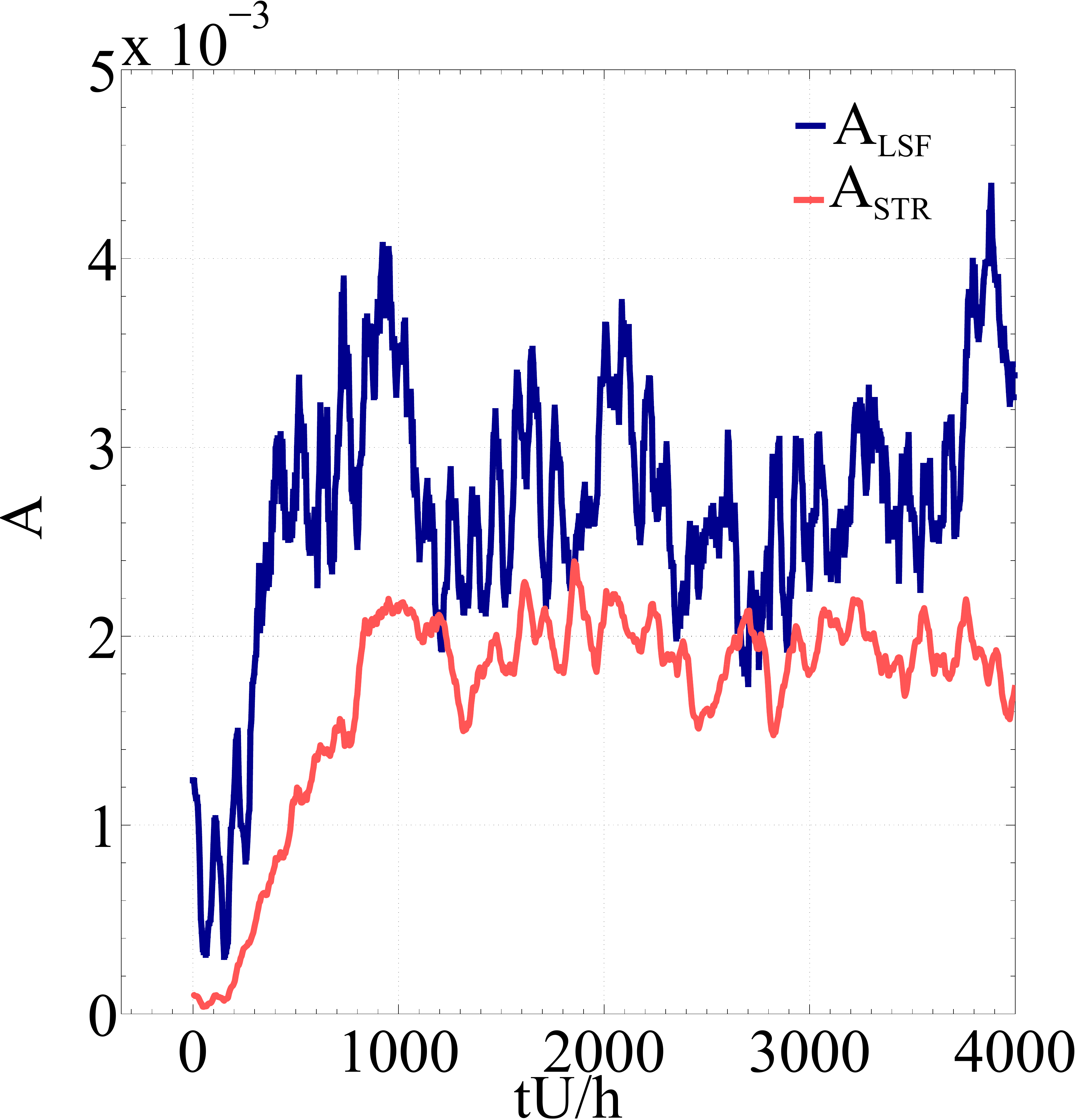}
	\\
		(c)
    \end{center}
	\end{minipage}
   	\begin{minipage}[c]{.32\linewidth}
	\begin{center}
\includegraphics[width=1\linewidth]{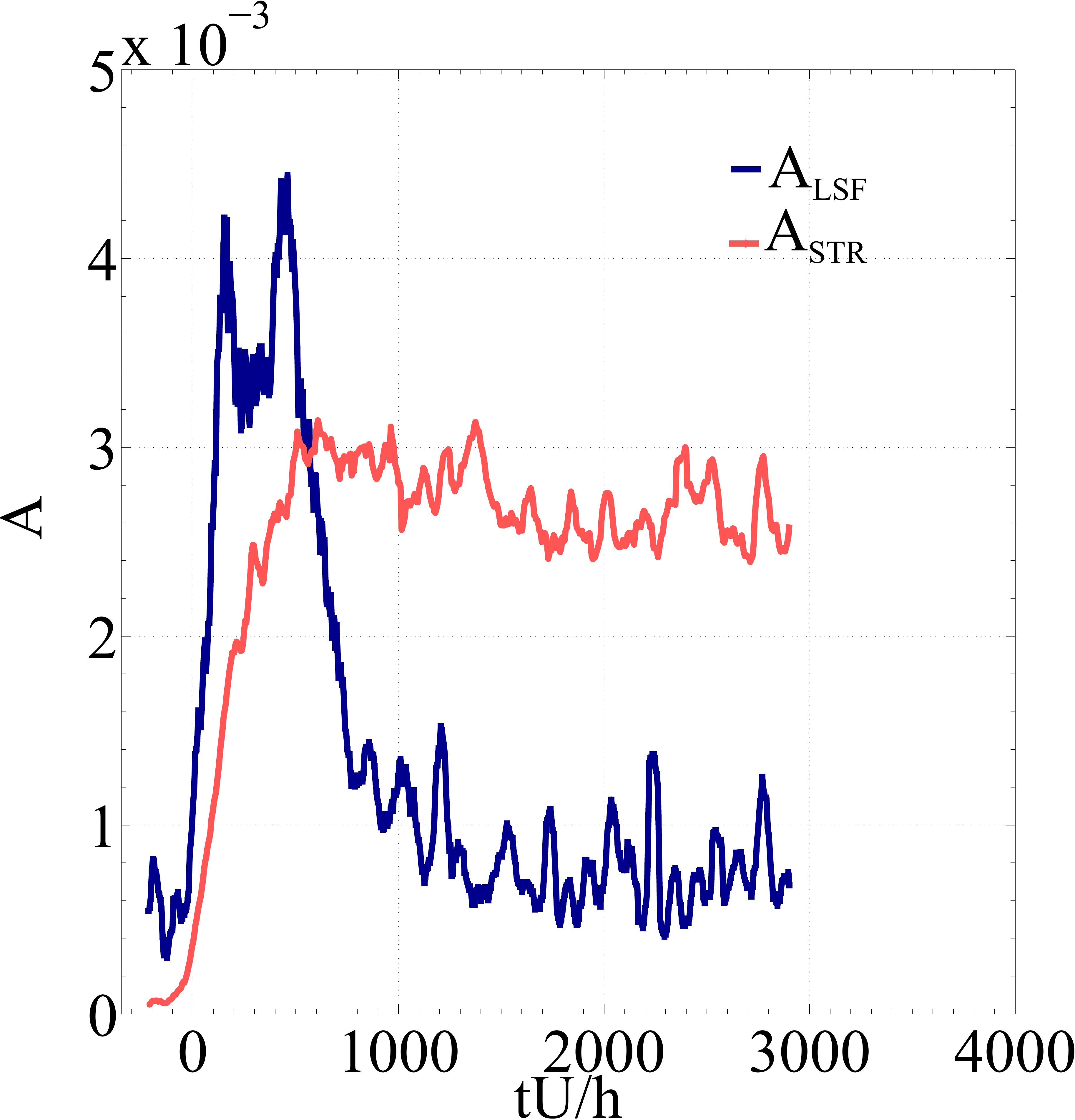}
	\\
   	(d)
   	\end{center}
   	\end{minipage}
\caption{Time evolution of streaks $A_{STR}$ (a) and large-scale $A_{LSF}$ (b) amplitudes for several Reynolds numbers. Time evolution of streaks $ A_{STR}$(red) and large-scale flows $A_{LSF}$(blue) amplitudes for given realizations at $Re=340$ (c) and $Re=403$ (d), $y/h\sim 0$.}
\label{max_ratio}
\end{figure}

\begin{figure}[h!]
\begin{minipage}[c]{.5\linewidth}
\includegraphics[width=0.9\linewidth]{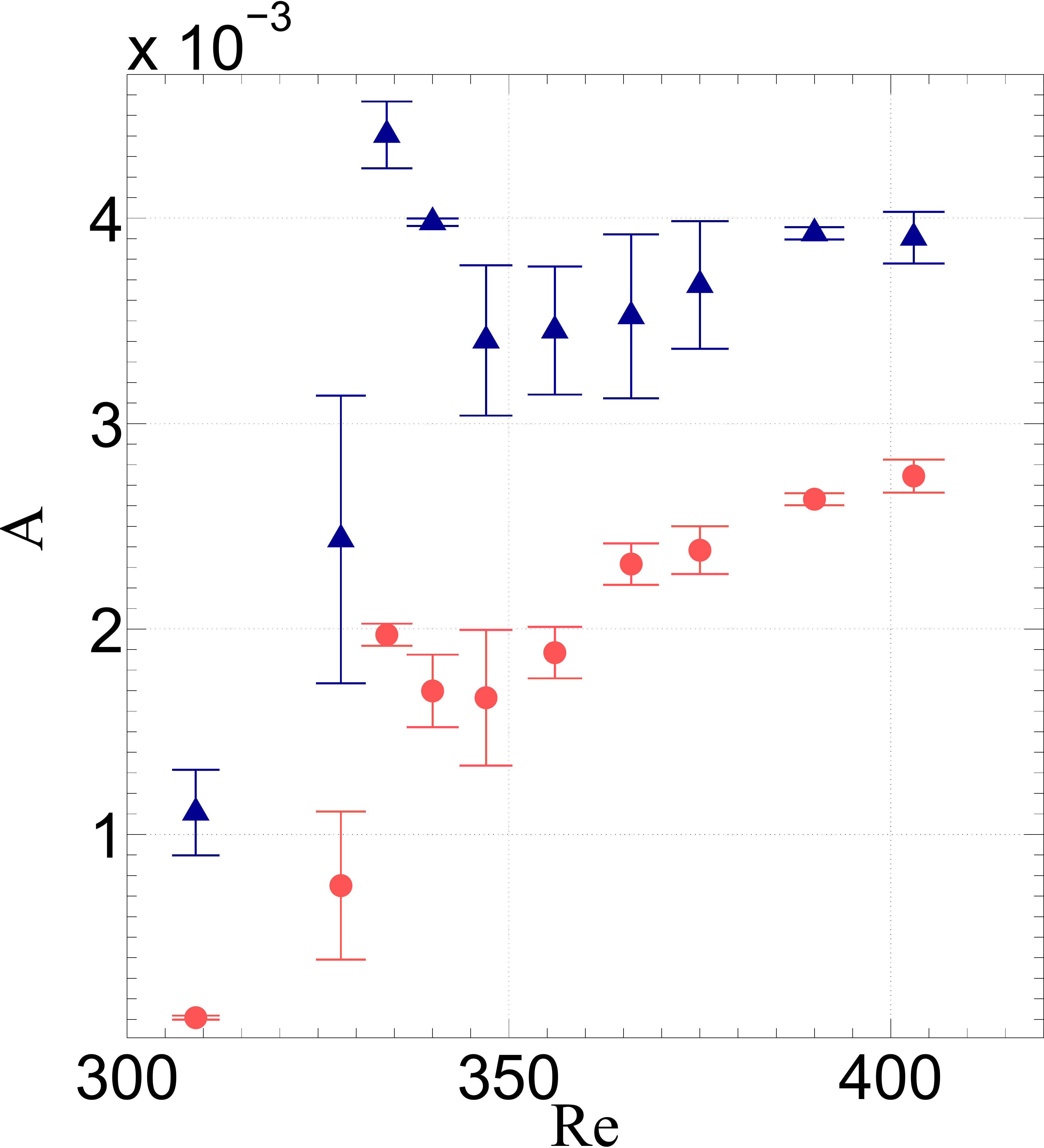}
\end{minipage}
\caption{Evolution with the Reynolds number of the maximal amplitude associated with the large scales (blue) and of the streaks plateau amplitude (red), $y/h\sim 0$.}
\label{max_ratio2}
\end{figure}

Fig.~\ref{fig:uramp} shows $U_z$ fields at different times along a step experiment during which a turbulent front invades the laminar flow from right to left in the pictures. In the first snapshot, the flow is laminar and the $U_z$ field is almost uniform. In the second one, a narrow turbulent area is already visible at the top right corner and a large scale recirculation very similar to that observed around the turbulent spot in fig.~\ref{fig:scalesep}-c spreads over half the picture. In the third snapshot, this large-scale structure has moved to the left along with the laminar-turbulent front. On the last snapshot the recirculation is present only close to the laminar-turbulent interface, while deep inside the turbulent area, far enough from the front, the velocity field is as in fig.~\ref{turb} where the flow is homogeneously turbulent. This exemplifies the fact that large-scale flows go along with the growth dynamics of turbulent areas. To better understand this interplay, we now consider the time evolution of the premultiplied power spectra.\\
Fig.~\ref{fig:Spectrogram_ez} show spectrograms associated with a step experiment (a) and a bead experiment (b). In both cases, the scenario is similar: during a first phase ($1100~h/U$ for the step, $300~h/U$ for the bead), spectra are almost uniform, do not present any peak and their magnitude are low: the flow is laminar. The noise visible at large scales defines the resolution of our PIV setup. During a second phase, energy emerges at large wave numbers (greater than $10h$), quickly followed by a third phase where energy raises around $\lambda / h \simeq 4-5$ while the energy at large scales still grows. This corresponds to the emergence of large-scale flows and to the establishment of well organised streamwise streaks ({\it i.e.} turbulence) on larger and larger areas. Note that the corresponding energy level are two to three times greater than the noise present in the first phase. In a final phase corresponding to a state where turbulence has invaded the whole flow, the large scales are damped and all the energy gathers around $\lambda /h\simeq 4-5$. We insist on the fact that the energy associated with large-scale structures appears earlier than the energy linked to turbulence. Spectra associated to five instants representing these three phases are extracted and plotted on fig.~\ref{fig:Spectrogram_ez}-c-d (see caption for details). On these plots, one better appreciates the low level of the noise during the laminar phase and the hierarchy of energy level of large and small-scales as time passes. Large scales are associated with the peak growing at $\lambda/h=40$. The amplitude of both peaks are similar even if the large-scale one is somehow higher.\\
We now focus only on the spectral components associated to large and small-scales. The small scale ({\it i.e.} the streaks) component $A_{STR}$ is defined as the average of $kE_z$ over $4 \leq \lambda/h \leq 5$ and the large-scale component $A_{LSF}$ as the value of $kE_z$ at $\lambda/h=40$. Their time evolutions are presented on fig.~\ref{max_ratio}-a and fig.~\ref{max_ratio}-b respectively. Each curve corresponds to a phase averaging of five realisations at a given value of $Re$. Before performing the phase average, a time sliding average over $1$ second ($5$ samples) is applied. Regarding streaks, the amplitude $A_{STR}$ remains roughly constant at a very low level before suddenly increasing during about $800~h/U$ to reach a well defined plateau. The level of this plateau (shown as horizontal lines in the figure) increases with the Reynolds number. %
The amplitude $A_{LSF}$ time evolution differs more from one value of $Re$ to another but three tendencies still appear. (i) For $Re\lessapprox 330$ $A_{LSF}$ increases monotonously to reach a plateau at relatively low level: large-scale flows are present but the growth is limited and the turbulent spot will not invade the whole spanwise extent of the experiment. These spots survive rather than grow. By contrast, if $Re \geq 330$, $A_{LSF}$ increases rapidly to reach a maximum. Then it either remains around this maximum (ii) or relaxes until stabilising on a more or less well defined plateau (iii). Distinction between (ii) and (iii) corresponds to spots eventually hitting the spanwise boundaries ($Re\geq350$) or not. Case (iii) includes situations with $Re\geq Re_t$ for which $A_{LSF}$ vanishes rather than relaxes. Note that the phase averaging may hide an important dispersion between various realisations obtained at the same $Re$. 
%
%
%
Cases (ii) and (iii) are better understood in fig.~\ref{max_ratio}-c and \ref{max_ratio}-d where large and small-scale amplitudes are plotted together for two given realizations at $Re=340$ and $Re=403$ respectively. For $Re=340$, from $t=0$, large-scale amplitude starts increasing before streaks amplitude does. Then both grow together before reaching plateaus. For $Re=403$, at a time around $10~h/U$, both amplitudes start to increase but this increase is faster for the large-scale than for the streaks. $A_{LSF}$ stabilizes to its maximal value while the streaks one keeps increasing. Once $A_{STR}$ reaches its plateau, $ A_{LSF}$ drops sharply to eventually vanish. This corresponds to situations above $Re_t$ for which the ultimate state is homogeneously turbulent and large-scale flows are thus transient.  Note that an overshoot is present at large scale. For range of $Re$ below $Re_t$ but above $Re\simeq350$) the spot develops until it reaches the spanwise edges of the experiment. Then, the turbulent fraction remains almost steady but the turbulent spot reorganises as oblique bands. Large-scale flows are present all along this reorganisation (the plateau mentioned above) but also once the bands are stabilised as shown numerically by Duguet and Schlatter \cite{duguet13_PRL}.\\
Fig.~\ref{max_ratio2} shows the evolution with the Reynolds number of the maximal amplitude associated to the large scales (defined as the maximum of the curves presented in fig.~\ref{max_ratio}-a and of the streaks amplitude (defined as the value of the plateau in fig.~\ref{max_ratio}-b. Error bars displayed in this figure correspond to the dispersion between various experiments performed at the same $Re$ we alluded to above. If the streaks amplitude is increasing smoothly in the range of $Re$ considered, a cross-over between two regimes may nevertheless be identified around $Re\simeq350$. At these values of $Re$, the behaviour of the large-scale maximal amplitudes changes abruptly from steep to moderate increase. All this would discriminate between the case (ii) and (iii) described above. This point will be further discussed in the next section.
%
%
\section{\label{sec:disc}Discussion and conclusions}
PIV measurements have been performed in plane Couette flow around laminar-turbulent interfaces associated to growing turbulent patches. Two very different protocols have been used, one with a localised perturbation achieved by a bead placed in the center of the channel and one with no specific perturbation but with turbulence generated at entrance denominated as step. In both cases, turbulence grows provided $Re$ is higher than $Re_g$. In the bead case, turbulence consist of a localised diamond like spot growing around the bead while in the step case turbulence arises from the stream-wise edges as large arrow shaped fronts. Spatial spectral analysis has been used to evidence for the first time the existence of large scale recirculations at the laminar-turbulent interface in an experimental plane Couette flow. Existence of large-scale flows in situations where laminar and turbulent domains coexist has already been explained by flow rate conservation arguments in presence of the overhang existing at the laminar-turbulent boundary \cite{lundbladh91_JFM,duguet13_PRL} or through additional pressure fields induced by the distribution of Reynolds stresses \cite{lagha07_POF}.

As soon as laminar and turbulent areas coexist in the experiment, even if turbulence eventually decays, large scale flows develops not only around the fronts but also far into the laminar and turbulent domains. Around the localised spots, they consist of quadrupolar recirculation cells as already reported in numerical simulations of plane Couette flow \cite{lagha07_POF,duguet13_PRL} and idealised shear flow \cite{schumacher01_PRE} but also experimentally in plane Poiseuille \cite{lemoult13_JFM}. Quantitative comparison to previous numerical works is difficult since the authors did not provide actual values to estimate the large-scale flow intensities. Nevertheless, we have performed numerically a similar but less exhaustive study on ChannelFlow \cite{channelflow,GibsonHalcrowCvitanovicJFM08} where we initiate spot growth with the initial perturbation used in \cite{lundbladh91_JFM}. The results obtained so far are in good quantitative and qualitative agreement with the present work, whenever the intrinsic differences between experimental and numerical setup do not prevent such comparison. Regarding our work, large-scale flow intensities are similar between the two types of experiments we have considered but they strongly depend on the Reynolds number as evidenced in fig.~\ref{max_ratio2} and on the turbulent fraction. This dependence on the turbulent fraction can be understood in the framework of flow rate conservation argument evoked above: the larger the turbulent fraction, the longer the overhang and thus the larger the flow rate imbalance that has to be compensated by stronger large scale flows.\\
According to our understanding and experimental observations, large-scale flows exist all along the growth process till turbulence has spread over the whole spanwise direction. Then two scenarios are possible: if $Re<330$, large-scale flow exist but spots survive rather than grow. If spots ultimately grow ({\it i.e.} if $Re\geq 330$) large-scale flows show rapid increase and eventually stabilise on finite value associated to a steady turbulent fraction. Our results suggest an additional distinction between experiments performed at $Re$ below or above $Re\simeq350$. In the latter case, triggered spots reach the spanwise extend of the experiment before reorganising as steady regular patterns. In the former, the steady ultimate turbulent fraction is too small to accommodate such patterns and the triggered spot starts reorganising as inclined patterns before reaching the spanwise edges. The final state is a wandering mix of inclined fractioned stripes whose turbulent fraction highly fluctuates around a stationary value. This \lq \lq critical\rq \rq \ $Re\simeq350$ is consistent with Prigent's work who reported that organised steady patterns can be observed only above $Re \simeq340$ while below stripes are not stable and one mostly observes stretched spots\cite{prigent01_phd}. Some preliminary numerical simulations similar to the present study confirm most of these observations and points out the resurgence of the large scales when the turbulent spot later reorganises to form laminar/turbulent stripes as those observed in large domains \cite{prigent02_PRL,barkley05_PRL,duguet10_JFM}.
An other intriguing and interesting feature is the role played by large-scale recirculations in the spreading of turbulence. On the one hand, we have shown that these recirculations are present as soon as laminar and turbulent domain coexist, growth being observed or not, but on the other hand, Schumacher and Eckhardt \cite{schumacher01_PRE} attest that no growth is observed when what they call the outflow is suppressed around a localised turbulent spot in a shear flow very similar to plane Couette. In a streamwisely constrained plane Couette geometry where no large scale flow is present, Duguet {\it et al.} \cite{duguet11_PRE} evidence a purely stochastic growth of turbulent spots but at rates far below what is observed in more realistic situations. Large-scale flows thus appear as a key ingredient of the growth even if they are not sufficient. Another point of debate is whether the large-scales are induced by or induce the growth. As seen in fig.~\ref{max_ratio}, large-scale amplitude increases before streaks amplitude does, implying that growth is initiated by these large-scale flows. How do they actually act on the growth, by simply advecting the turbulent region or by locally modifying the laminar profile to initiate a growth by destabilisation \cite{gadelhak81_JFM}, remains an open question. The energy transfers between large and small-scales evidenced in the present article should be scrutinised. We are currently studying the growth rates of turbulent spots in bead experiments through high resolution visualisations in an attempt to refine our knowledge of the spot growth mechanisms. We hope to be able to discriminate the role played by large-scale flows in this process.
\section{\label{sec:level5}Acknowledgements}
The authors would like to thank Y. Duguet, P. Manneville and O. Cadot for helpful discussions. The KTH and Saclay groups are acknowledged for sharing their experience regarding the experimental set-up, particularly F. Daviaud for lending the polygonal mirror. The master students L. Colo, G. Anyigba, E. Bach, C. Jacob, E. Sieffert, P. Yang and Y. Zhu are also acknowledged for their contributions. F. Moisy is acknowledged for is PIVMat toolbox. This project was supported by ANR Jeunes chercheurs/Jeunes chercheuses QANCOUET.

\end{document}